\documentclass[%
reprint,
superscriptaddress,
nofootinbib,
 amsmath,amssymb,
 aps,
]{revtex4-2}

\usepackage{graphicx}
\usepackage{dcolumn}
\usepackage{bm}
\usepackage{braket}
\usepackage{upgreek}
\usepackage{xcolor}
\usepackage{afterpage}
\usepackage{relsize}
\usepackage{multirow}
\usepackage{float}
\usepackage{physics}
\usepackage{makecell}
\usepackage{placeins}
\usepackage[colorlinks=true,urlcolor=blueprl,citecolor=blueprl,linkcolor=blueprl]{hyperref}
\usepackage{soul}
\usepackage{hhline}

\newcommand{\ZJ}[1]{\textcolor{black}{#1}}
\newcommand{\ozlem}[1]{\textcolor{black}{#1}}
\newcommand{\REVozlem}[1]{\textcolor{black}{#1}}

\newcolumntype{P}[1]{>{\centering\arraybackslash}p{#1}}

\usepackage{hyperref}
\let\svthefootnote\thefootnote
\newcommand\freefootnote[1]{%
  \let\thefootnote\relax%
  \footnotetext{#1}%
  \let\thefootnote\svthefootnote%
}

\definecolor{blueprl}{RGB}{46,48,146}

\def\ANU{Centre of Excellence for Quantum Computation and Communication Technology, The~Department~of~Quantum Science and Technology, Research School of Physics and Engineering, The Australian National University, Canberra, Australian Capital Territory, Australia}
\def\ASTAR{A*STAR Quantum Innovation Centre (Q.InC), Agency~for~Science,~Technology~and~Research~(A*STAR), 2 Fusionopolis Way, Innovis \#08-03, Singapore 138634, Republic of Singapore}

\def\CSIRO{Data61, Commonwealth Scientific and Industrial Research Organisation, Marsfield 2122, NSW, Australia}
\def\QLabs{Quintessence Labs, Canberra, Australia}

\begin{document}


\title{Enhanced continuous-variable quantum key distribution protocol via adaptive signal processing}
\author{\"{O}zlem Erk{\i}l{\i}\c{c}$^{*,\dagger}$}
\freefootnote{$^*$ \href{ozlemerkilic1995@gmail.com}{ozlemerkilic1995@gmail.com}}
\freefootnote{$^\dagger$ These authors contributed equally}
\affiliation{\ANU}

\author{Biveen Shajilal$^\dagger$}
\affiliation{\ANU}
\affiliation{\ASTAR}

\author{Lorc\'{a}n O. Conlon}
\affiliation{\ANU}
\affiliation{\ASTAR}

\author{Angus Walsh}
\affiliation{\ANU}
\author{Aritra~Das}
\affiliation{\ANU}
\author{Sebastian Kish}
\affiliation{\CSIRO}
\author{Thomas Symul}
\affiliation{\QLabs}
\author{Ping Koy Lam}
\affiliation{\ANU}
\affiliation{\ASTAR}
\author{Syed M. Assad}
\affiliation{\ANU}
\affiliation{\ASTAR}
\author{Jie Zhao}
\affiliation{\ANU}

\date{\today}
             
\maketitle

\section*{Abstract}
\ozlem{Quantum key distribution (QKD) provides a promising approach to secure communications,} with continuous-variable QKD (CV-QKD) offering compatibility with existing telecommunication infrastructure. Despite this advantage, CV-QKD is limited by challenges such as losses in terrestrial fibres and atmospheric scintillation in free-space channels. 
We introduce a QKD protocol that surpasses the optimal Gaussian modulated CV-QKD (GG02) protocol by utilising probabilistic filters without known physical representation. Our approach employs a Gaussian filter at Alice's station and a non-Gaussian notch-like filter at Bob's station. Alice’s filter optimises modulation variance to achieve key rates near the optimal GG02 performance, while Bob’s filter adapts the effective channel conditions, which can result in higher key rates than the optimal GG02 protocol. 
Our security analysis avoids Gaussian extremality, accurately bounding Eve’s information. The protocol dynamically optimises the secret-key rate for rapidly changing channels, such as terrestrial links and satellite-to-ground communications, and can extract keys in regions deemed non-secure by parameter estimation. Implemented at software level, our protocol requires no hardware modifications and can be integrated into existing \ZJ{QKD systems.} \ZJ{Experimental results show a threefold increase in key rates over the optimal GG02 protocol, while simulations for Low Earth Orbit satellite quantum communications indicate a 400-fold increase compared to the non-optimised counterpart.}

\section*{\label{sec:introduction}Introduction}
Quantum key distribution (QKD) is a method used to securely establish a secret key between two distant parties, Alice and Bob~\cite{ekert2014, gisin2002, pirandola2020advances}. One variant of QKD is continuous-variable (CV) protocols, where keys are encoded in the amplitude and phase quadratures of the optical field and measured using heterodyne or homodyne detection~\cite{ralph1999, hillery2000}. \ZJ{CV-QKD can be categorised into different types, one of which includes protocols with Gaussian modulation (GM), where coherent states are modulated with a Gaussian distribution~\cite{grosshans2002continuous, grosshans2003quantum, weedbrook2004quantum, lance2005no, lodewyck2007quantum, madsen2012continuous, jouguet2013experimental, wang201525, zhang2020long, jain2022practical, hajomer2024long}.} The most commonly used CV-QKD protocol with GM is the GG02 protocol~\cite{grosshans2002continuous, grosshans2003quantum, weedbrook2004quantum}, which optimises the variance of Gaussian modulation for each transmission distance to achieve the best key rates. A key step in QKD is parameter estimation, \ZJ{which is often performed after error reconciliation to assess channel noise and loss, helping to bound Eve's information~\cite{garcia2007quantum}}. This helps optimise experimental parameters, assuming the channel varies slowly. However, if parameter estimation falls within a non-secure region, key data must be discarded, and rapid variations like atmospheric scintillation can render these estimates ineffective.

\begin{figure*}[htp!]
\center{\includegraphics[width=\textwidth]
        {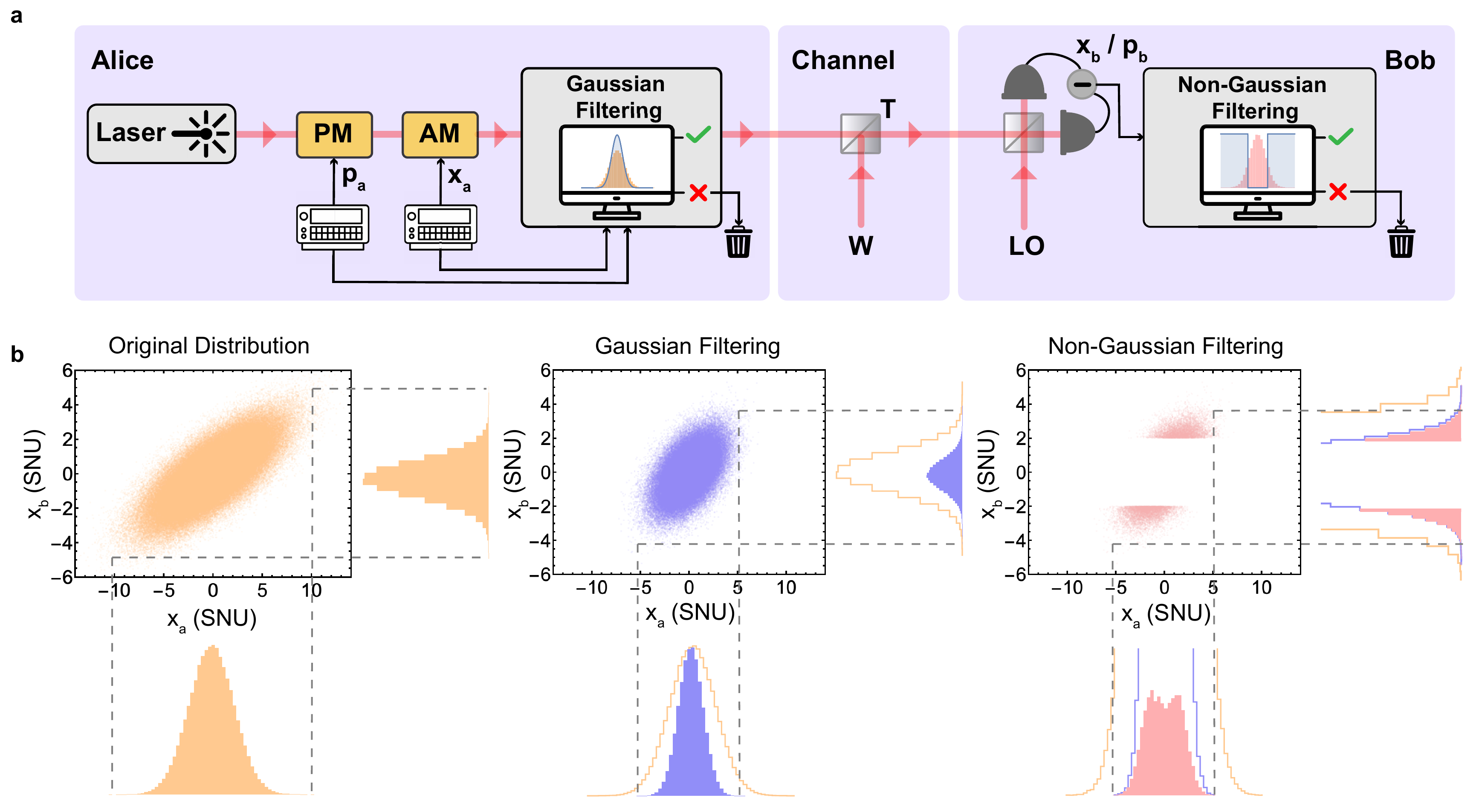}}
\caption{\label{fig:experimental_setup} Experimental schematic and illustrative diagram of the filtering process. \textbf{(a)} Alice encodes coherent states $\ket{(x_a+ip_a)/2}$ from a Gausssian distribution with zero mean and variance, $V_{mod}$, using white noise generated by function generators. She samples $x_a$ and $p_a$ simultaneously, then sends the encoded states through a quantum channel with transmittivity $T$ and thermal noise $W$. Bob performs homodyne detection to measure $x_b$ and $p_b$, randomly switching between these two quadratures. After data acquisition and once the states have been sent to Bob, Alice applies a Gaussian filter to the measurement data kept on her computer, followed by Bob applying a non-Gaussian filter to his measurement outcomes. \textbf{(b)} Left: The correlation between Alice's variables $x_a$ and Bob's measurement outcomes $x_b$ before any post-selection. Middle: The correlation after Alice's Gaussian post-selection \textcolor{black}{superimposed by the original distribution (orange)}, which reduces the variance of the Gaussian-modulated variables, effectively creating a thermal state with this \ozlem{reduced} variance. Right: The correlations after Bob's non-Gaussian post-selection, resulting in highly non-Gaussian distributions for $x_a$ and $x_b$. PM/AM: Electro-optic phase/amplitude modulators and LO: Local oscillator. 
}
\end{figure*}
\ZJ{In CV-QKD, passive methods that do not require feedback loops to optimise experimental parameters can enhance key rates by employing quantum processes such as noiseless linear amplifiers (NLAs)~\cite{xiang2010heralded, winnel2020generalized}, phase-sensitive amplifiers~\cite{bencheikh2001quantum}, photon subtraction~\cite{huang2013performance, hu2020continuous}, and photon catalysis~\cite{ye2019improvement, hu2020continuous, hu2021performance}. While these techniques can increase key rates, they are challenging to implement.} \ZJ{Recent post-selection protocols emulate such physical processes using digital filtering, where applying a quantum process becomes equivalent to first detecting the quantum state and then digitally filtering the measured outcomes.} This approach is particularly useful for point-to-point applications like QKD.
\ZJ{For example, measurement-based NLA (MB-NLA) replaces physical NLAs by employing Gaussian post-selection on detection outcomes.} \ZJ{This approach extends the transmission distance of CV-QKD~\cite{walk2013security, fiuravsek2012gaussian, hosseinidehaj2020finite} and also provides overall performance enhancements in CV-based quantum communications~\cite{chrzanowski2014measurement, zhao2017characterization, zhao2023enhancing, shajilal2024improving}.} Similar approaches have been applied to photon subtraction and photon catalysis protocols, where physical processes are replaced with non-Gaussian virtual photon-subtraction~\cite{li2016non, zhong2018self} and Gaussian zero-photon catalysis post-selection~\cite{zhong2020virtual}. \ZJ{However, since these methods still emulate physical processes, the post-selection filters are constrained by the characteristics of those processes.} Additionally, these protocols rely on Gaussian extremality, which states that Gaussian states maximise the von Neumann entropy for a given covariance matrix~\cite{garcia2006unconditional, wolf2006extremality}. This results in an overestimation of Eve's information with non-Gaussian filters, leading to an underestimation of the key rate.

In this work, we present a new CV-QKD protocol using Gaussian modulations and homodyne or heterodyne detection followed by post-selection. The protocol introduces a paradigm shift by eliminating the need to simulate physical processes for filtering. It combines post-selection at both Alice’s and Bob’s stations: Alice's Gaussian filter optimises modulation variance to achieve key rates near the optimal GG02 performance, while Bob's non-Gaussian filter emulates a higher-performing channel, surpassing optimal GG02 key rates. To avoid overestimating Eve's information, we developed a security analysis that tightly bounds Eve’s information after Bob’s filter. The protocol dynamically optimises key rates for rapidly changing channels, such as satellite-to-ground links, where rapid channel changes render any parameter estimation unreliable. Moreover, it can extract keys even in regions where unfiltered channel parameters would not allow secure QKD under the GG02 protocol.

\section*{\label{sec:protocol}Protocol}
\begin{figure*}[htp!]
\center{\includegraphics[width=\textwidth]
        {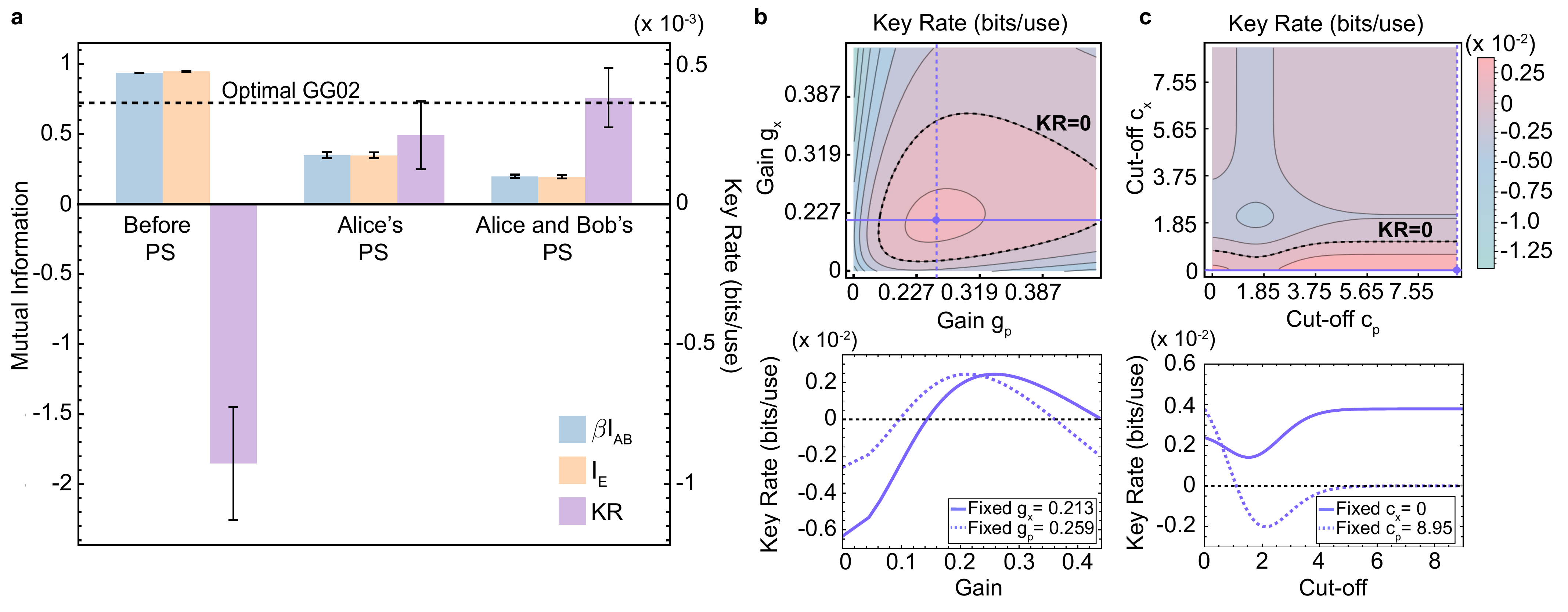}}
\caption{\label{fig:figure2} Experimental results of a quantum channel with \ozlem{a transmission of $T=0.26$, equivalent to a fibre distance of 29.1 km (assuming 0.2 dB loss per km),} with a thermal noise of $W_x=1.02\pm0.001$ and $W_p=1.01\pm0.001$ in shot noise units (SNU) in $x$ and $p$ quadratures, respectively. \textbf{(a)} From left to right, we give the mutual information between Alice and Bob (blue boxes), Eve’s information (orange boxes) and the key rates of the protocol (purple boxes) before any post-selection, after Alice’s post-selection and after Alice and Bob’s post-selection combined, respectively. \ozlem{The key rate is initially negative, indicating that Eve’s information exceeds the mutual information of Alice and Bob, making it a non-secure region. After Alice’s post-selection, the key rate becomes positive but remains below the optimal GG02 rate for that distance.} The key rate is further optimised by applying a post-selection on Bob’s side where the key rate achieves the optimal GG02 line. \textbf{(b)} The top figure shows the contour plot of the key rates after Alice's post-selection along the $x$ and $p$ quadratures, with gains ranging from 0 to 0.436. The bottom figure illustrates the key rate slices at optimal performance after Alice's post-selection with fixed gains $g_x$ and $g_p$. The solid line represents fixed $g_x$ while varying $g_p$, and the dashed line represents fixed $g_p$ while varying $g_x$. \textbf{(c)} Top figure demonstrates the contour plot of the key-rates after Bob’s post-selection with the cut-off parameters ranging from 0 to 8.95, while the bottom figure shows the key rates after Bob's post-selection with fixed optimal cut-offs in the $x$ (solid line) and $p$ (dashed line) quadratures. The solid line varies the cut-off in the $p$ quadrature while keeping the $x$ cut-off fixed, and the dashed line varies the cut-off in the $x$ quadrature while keeping the $p$ cut-off fixed. KR: Key rate, $\mathrm{I_{AB}}$: Mutual information between Alice and Bob, $\mathrm{I_E}$: Eve's information and $\beta$: Reconciliation efficiency. The reconciliation efficiency is set to $92\%$ \ozlem{throughout the experiment.} The error bars indicate the statistical uncertainties in the data. For details on how they are calculated, refer to the Methods section.
}
\end{figure*}
The experimental setup is shown in Fig.~\ref{fig:experimental_setup}(a), where Alice generates two variables, $x_a$ and $p_a$, from a zero-mean Gaussian distribution with variance $V_{mod}$ in units of vacuum noise using two independent function generators. In our implementation, Alice’s Gaussian-modulated variables are encoded onto the sidebands of a 1064~nm continuous-wave laser using amplitude and phase electro-optic modulators driven at 3–5~MHz. This generates a thermal state with quadrature variance $V_{mod}+1$, as required by the protocol. Alice retains these variables to apply a Gaussian filter to both $x_a$ and $p_a$ after the states have been sent to Bob and measured. The encoded state is sent through a quantum channel with transmittance $T$ and thermal noise $W$, simulated by a half-wave plate and polarising beamsplitter. Bob performs homodyne detection, randomly measuring the $x$ or $p$ quadrature. The homodyne signal is filtered by a $3-5$ MHz bandpass filter and then fed to the oscilloscope with a sampling rate of $25$ MSamples$/$s, collecting $10$ million samples. Both homodyne data and Alice's data were digitally filtered with a $3-3.5$ MHz bandpass filter.

\begin{figure*}[htp!]
\center{\includegraphics[scale=0.4]
        {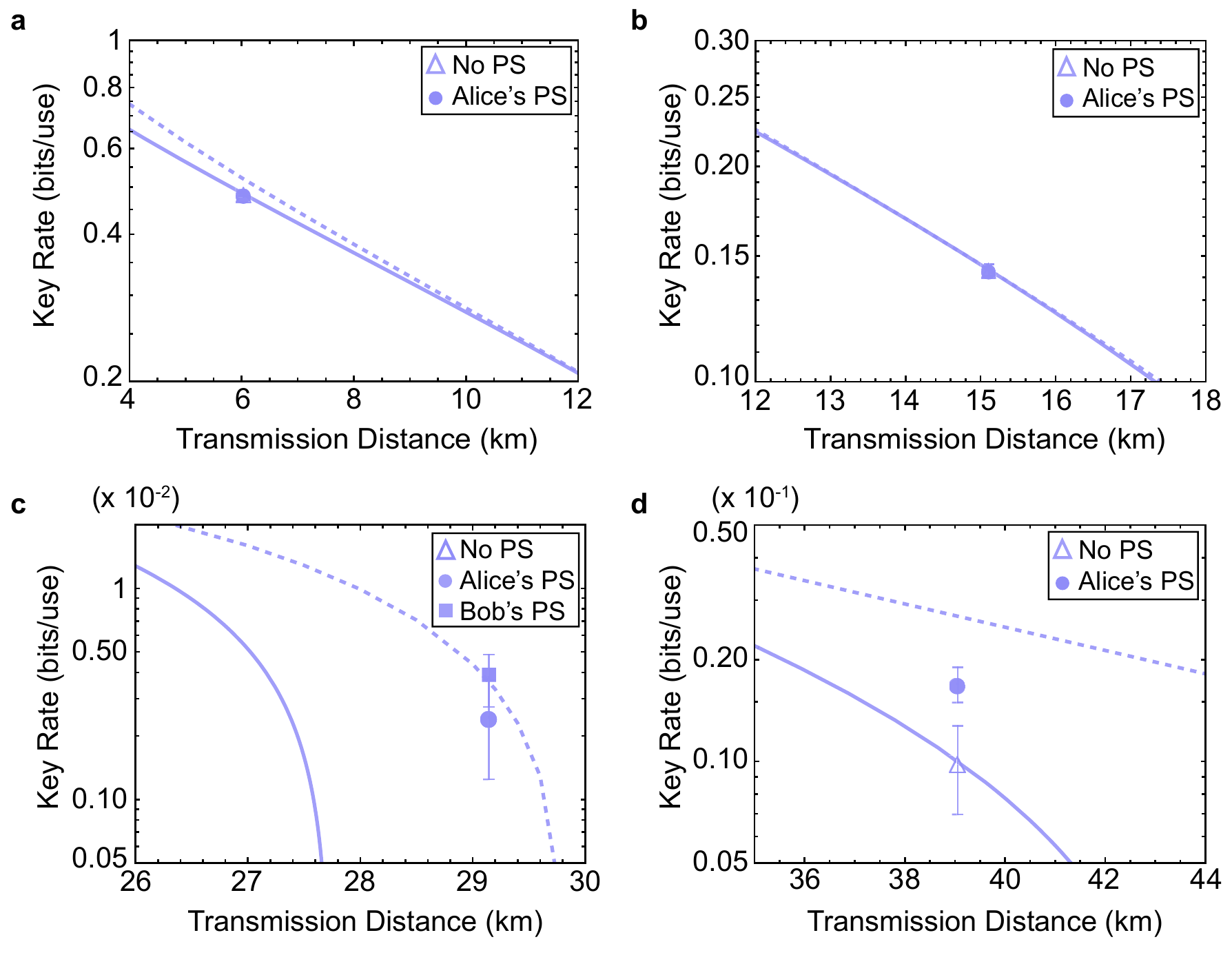}}
\caption{\label{fig:figure3} Experimental key rates for various channel parameters. Solid lines represent theoretical key rates of the GG02 protocol with fixed modulation variance, while dashed lines show theoretical key rates with optimal variances. Triangle markers indicate raw key rates before post-selection, circles show key rates after Alice's post-selection, and the square marker represents results after Bob's post-selection. \textbf{(a)} and \textbf{(b)} show the experimental results at $d=6$ km \ozlem{($T=0.76$)} and $d=15.1$ km \ozlem{($T=0.50$)}, respectively. For $d=6$ km, thermal noise is $W_x=1.02\pm0.004$~SNU and $W_p=1.04\pm0.004$~SNU; for $d=15.1$ km, $W_x=1.02\pm0.002$~SNU and $W_p=1.03\pm0.002$~SNU. Key rates before and after Alice's post-selection overlap, indicating optimal operations. \textbf{(c)} Key rates at $d=29.1$ km \ozlem{($T=0.26$)} with thermal noise $W_x=1.02\pm0.001$~SNU and $W_p=1.01\pm0.001$~SNU. The post-selections enhance the key rate which is otherwise negative; \ozlem{as a result, the data without post-selection (triangular marker) does not appear on the logarithmic scale.} \textbf{(d)} Key rates at $d=39.1$ km \ozlem{($T=0.17$)} with thermal noise $W_x=1.00\pm0.001$~SNU and $W_p=1.00\pm0.001$~SNU, where only Alice's post-selection was applied, as Bob's post-selection did not lead to any further improvements in this region.} 
\end{figure*}
While the experimental setup follows the well-known GG02 protocol~\cite{grosshans2002continuous,grosshans2003quantum}, which utilises Gaussian modulated coherent states and homodyne detection, our protocol diverges at the software level. This software-level protocol can be applied to any system with Gaussian modulated states and coherent detection. After the data acquisition, Alice's post-selection is applied on the samples $x_a$ and $p_a$ where Alice's Gaussian post-selection filter is defined by the following expression
\begin{equation}
    \label{eq:alices_filter}
    F_A(x_a)=e^{-g_x^2 x_a^2},
\end{equation}
where  $x_a$ is the randomly generated variable and  \text{$g_x\geq0$} represents the filter gain. Each symbol, $x_a$, is either kept or discarded based on the probability $F_A(x_a)$ (See the Methods section for details on how the filter is applied). The same function applies to $p_a$, replacing $g_x$ with $g_p$.  This post-selection on $x_a$ and $p_a$ imitates attenuating the modulation variance, as shown in Fig.~\ref{fig:experimental_setup}(b). When $g_x=0$, no filtering is applied, but as $g_x$ increases significantly (e.g., $g_x = 10^5$), the modulation variance approaches zero, effectively turning the state into vacuum. In the GG02 protocol, there is an optimal modulation variance for each distance and noise combination to achieve the best key rates. 
While Alice's post-selection does not alter the channel parameters, it reduces the modulation variance to match the optimal GG02. Unlike GG02, which requires modulation variance optimisation, Alice’s approach optimises both the variance and key rates without needing prior channel characterisation. The optimisation also inherently depends on the reconciliation efficiency, $\beta$, which quantifies how efficiently Alice and Bob can extract shared bits from their correlated data during classical post-processing (see Eq.~\eqref{eq:keyrate_equation}). Lower values of $\beta$ reduce the mutual information between Alice and Bob, shifting the modulation variance that maximises the final key rate. Therefore, when comparing key rates or optimising filters, the effect of $\beta$ must be taken into account.

The key rates can be further optimised by using a non-Gaussian post-selection performed by Bob. Bob's post-selection filter can be expressed as
\begin{equation}
    F_B(x_{b}) =
    \begin{cases}
	0 & \text{$-c_x<x_{b}<c_x$} \\ 
	1 & \text{$\mathrm{elsewhere}$},
    \label{eq:bobs_filter}
    \end{cases}
\end{equation}
where $x_b$ represents Bob's homodyne measurement outcome and $c_x$ is the cut-off parameter of the filter which can take values from $0$ to the largest measurement outcome of Bob. Bob's post-selection discards all the numbers within the cut-off, therefore, it creates a highly non-Gaussian distribution as shown in Fig.~\ref{fig:experimental_setup}(b). This post-selection effectively simulates a channel with better performance than the original one used by Alice to send the quantum states.

In QKD, Eve holds the purification of Alice and Bob's joint state $\sigma_{AB}$, meaning her von Neumann entropy matches that of the state shared by Alice and Bob. Therefore, Eve's information can be bounded by the Holevo bound~\cite{holevo1998capacity}, which represents the maximum classical information extractable from a quantum channel. \ozlem{It is expressed as}
\begin{equation}
    I_E=S(\sigma_{AB})-S(\sigma_{A|B}),
    \label{eq:eves_information_purification}
\end{equation}
where $S(\sigma_{AB})$ and $S(\sigma_{A|B})$ are the von Neumann entropies of Alice and Bob's joint state and Alice's state conditioned on Bob's measurement outcome $x_b$, respectively. Since Bob's post-selection is non-Gaussian, using the covariance matrices $\sigma_{AB}$ and $\sigma_{A|B}$ would overestimate Eve's information due to Gaussian extremality, leading to lower key rates. Instead, we directly calculate Eve's information from her density matrix, assuming that she performs a general Gaussian attack. In this attack, Eve prepares a two-mode squeezed vacuum (TMSV) state with a variance that mimics the quantum channel~\cite{weedbrook2012gaussian}. She replaces the channel between Alice and Bob with a beamsplitter, injecting one mode ($E_1$) into the beamsplitter to interact with Alice's state, while retaining the other mode ($E_2$)~\cite{weedbrook2012continuous}. 
While the optimal strategy after post-selection is not known, our assumption follows from the intuition that Alice and Bob can verify (within some tolerance) that the channel remains Gaussian by computing higher-order moments from their measurement outcomes such as skewness and kurtosis, thereby restricting Eve to Gaussian attacks. 
Eve's information is then calculated using the Holevo bound as follows
\begin{equation}
\label{eq:eves_information_density_protocol}
    I_{E}=S(\rho_{E_1E_2})-S(\rho_{E_{1}E_{2}|B}(x_b)),
\end{equation}
\begin{table*}[hbt!]
\caption{\label{tab:all_experimental_parameters}
Experimental parameters of the state-of-the-art CV-QKD demonstrations. 
}
\begin{ruledtabular}
\begin{tabular}{c|c|c|c|c|c|c|c|c|c|c|c}
\shortstack{\vspace{2mm}Experiment\vspace{2mm}} & 
\shortstack{\vspace{2mm}Detection\vspace{2mm}} & 
\shortstack{\vspace{1mm}$V_{mod}$ \\ (SNU)\vspace{1mm}} &
\shortstack{\vspace{1mm}Distance \\ (km)\vspace{1mm}} &
\shortstack{\vspace{1mm}Loss \\ (dB)\vspace{1mm}} &
\shortstack{\vspace{1mm}$\xi_{ch}$ \\ (SNU)\vspace{1mm}} &
\shortstack{\vspace{1mm}$\xi_{d}$ \\ (SNU)\vspace{1mm}} &
\shortstack{\vspace{1mm}$\xi_{t}$ \\ (SNU)\vspace{1mm}} &
\shortstack{\vspace{1mm}$\eta$ \\ (\%)\vspace{1mm}} &
\shortstack{\vspace{1mm}$\beta$ \\ (\%)\vspace{1mm}} &
\shortstack{Key Rate \\ Improvement \\ (\%)\footnote{Key rate improvements are calculated at the experimental distances (Fig.~\ref{fig:figure4}).}} &
\shortstack{Distance \\ Improvement\\(km)\footnote{Distance improvements are estimated by extrapolating performance beyond the experimental range (Fig.~\ref{fig:figure5}). As such, Figs.~\ref{fig:figure4} and ~\ref{fig:figure5} are evaluated under different conditions.}}\\
\colrule
Wang \textit{et al.}~\cite{wang201525} & Homodyne & 15 & 50 & 10 & 0.07 & 0.05 & - & 60.141 & 96.7 & 11.3 & 42\\ \hline
\rule{0pt}{3ex}
\shortstack{Hajomer \textit{et al.}~\cite{hajomer2024long}\vspace{0.05mm}} & \shortstack{Heterodyne\vspace{0.05mm}} & \shortstack{8.41\vspace{0.5mm}} & \shortstack{100\vspace{0.5mm}} & \shortstack{15.52\vspace{0.5mm}} & \shortstack{$0.212\times10^{-3}$\vspace{0.5mm}} & \shortstack{0.06272\vspace{0.5mm}} & \shortstack{-\vspace{0.5mm}} & \shortstack{68\vspace{0.5mm}} & \shortstack{92.5\vspace{0.5mm}} & \shortstack{16.4\vspace{0.5mm}} & \shortstack{19\vspace{0.5mm}}\\ \hline
\rule{0pt}{4ex}\shortstack{Data61 - CSIRO\vspace{1mm}\footnote{Experimental imperfections cause asymmetry in the $x$ and $p$ quadratures. The first row shows the experimental values for $x$, while the second row shows the values for $p$. $V_{mod}$: Modulation variance in shot noise units (SNU), $\xi_{ch}$: Channel noise, $\xi_{ch}$: Dark noise, $\xi_{t}$: Trusted noise attributed to state preparation, $\eta$: \textcolor{black}{Detection} efficiency, $\beta$: Reconciliation efficiency.}} & \shortstack{Heterodyne\vspace{0.7mm}} & \shortstack{4.06 \\ 4.21\vspace{-0.3mm}} & \shortstack{41.71\vspace{1mm}} & \shortstack{7.92\vspace{1mm}} & \shortstack{0.0437 \\ 0.0654\vspace{-0.3mm}} & \shortstack{0.4401 \\ 0.4459\vspace{-0.3mm}} & \shortstack{0.0867 \\ 0.0933\vspace{-0.3mm}} & \shortstack{20.6\vspace{1mm}} & \shortstack{92.5\vspace{1mm}} & \shortstack{206.2\vspace{1mm}} & \shortstack{1\vspace{1mm}} \\ \hline
Zhang \textit{et al.}~\cite{zhang2020long} & Homodyne & 14.23 & 140.52 & 23.46 & 0.0219 & 0.2717 & - & 61.34 & 96 & 74.5 & 42 \\ \hline
Li \textit{et al.}~\cite{li2023continuous} & Heterodyne & 10 & 100 & 18.96 & 0.0692 & 0.18 & - & 42 & 96.7 & 401.5 & 23 
\end{tabular}
\end{ruledtabular}
\end{table*}
\begin{figure*}[hbt!]
\center{\includegraphics[scale=0.33]
        {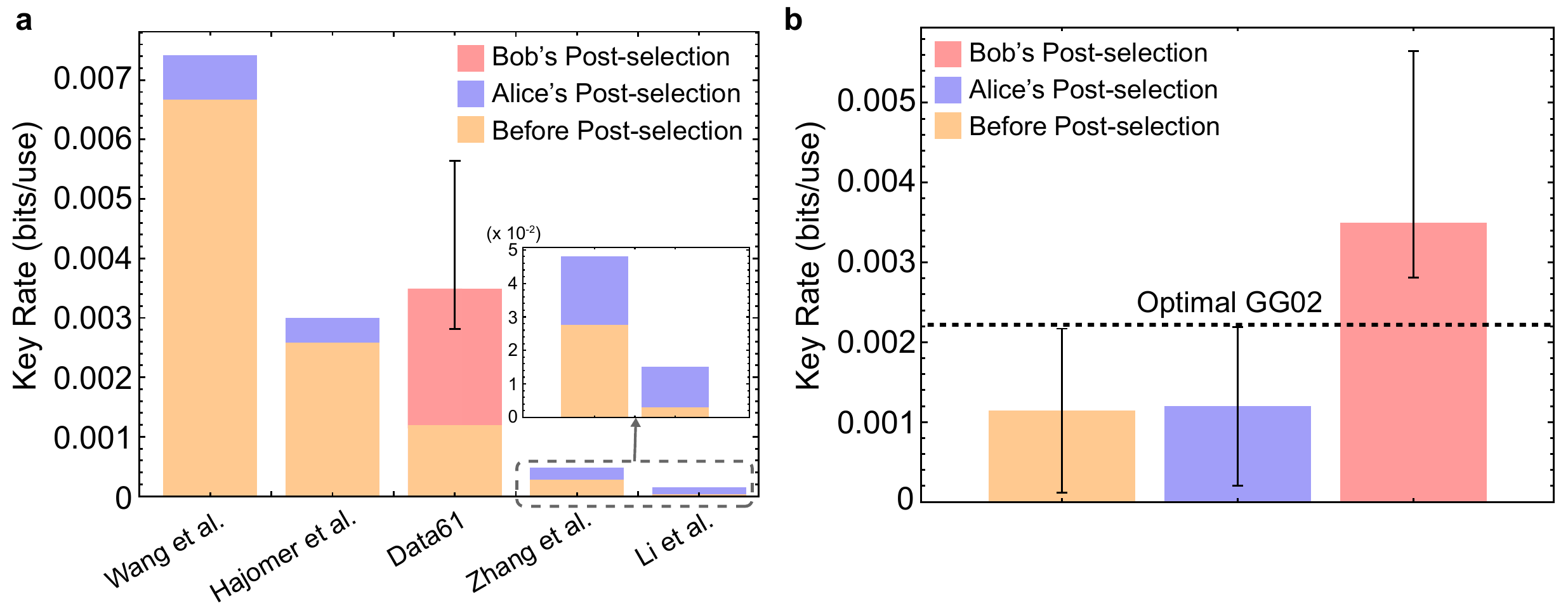}}
\caption{\label{fig:figure4} Key rates of existing experiments after applying our protocol, based on their original experimental parameters which are shown in \ozlem{Table}~\ref{tab:all_experimental_parameters}. \textbf{(a)} Key rates from various experiments~\cite{wang201525,hajomer2024long,zhang2020long,li2023continuous} and the experimental data from Data61 using the CV-QKD setup from Quintessence Labs. Orange boxes represent the key rates before post-selection, purple boxes denote the results after Alice's post-selection, and the pink box shows the key rate after Bob's post-selection. \textbf{(b)} Key rates of Data61 at each step, with a reconciliation efficiency of $\beta=92.5\%$ \ZJ{(Refer to Supplementary Information II for the derivation of the key rate with heterodyne detection)}. \ZJ{The error bars for Bob's post-selection are based on worst and best-case estimates.}}
\end{figure*}
where $S(\rho_{E_1E_2})$ represents the von Neumann entropy of Eve's average state, and $S(\rho_{E_{1}E_{2}|B})$ gives the von Neumann entropy of Eve's conditional state given Bob measures an outcome $x_b$. \textcolor{black}{This equation saturates Eq.~\eqref{eq:eves_information_purification} when Bob performs no post-selection, and everything remains Gaussian. In that case, Eve's average state is the sum of all Gaussian conditional states measured by Bob. With non-Gaussian post-selection, the conditional states remain Gaussian, but Eve's average state becomes non-Gaussian. Since Bob can discard measurement outcomes within the filter cut-off, Eve's average state is no longer obtained by simply summing all the states, but is derived from the filtered results as follows}
\begin{equation}
    \rho_{E_1E_2}=\sum\limits_{x_{i}=-n}^{n} p_b'(x_i) \rho_{E_{1}E_{2}|B}(x_{i}),
    \label{eq:eve_ave_denmat}
\end{equation}
where $n$ is the maximum number that Bob measures and $p_b'(x_i)$ denotes Bob’s probability of getting a given outcome, $x_i$ (For a detailed derivation of how to obtain the density matrices $\rho_{E_1E_2}$ and $\rho_{E_{1}E_{2}|B}$, refer to the Methods section). This method is particularly beneficial because it does not rely on Gaussian extremality to determine key rates. Previously, this reliance limited the performance of some protocols when non-Gaussian post-selection was used~\cite{li2016non}.

It is important to note that post-selection by both parties is not always required. Depending on the channel parameters, either Alice's post-selection or Bob's post-selection alone may be sufficient, while in some cases, both may be necessary. The advantage of our protocol lies in its ability to \ZJ{be augmented into} existing CV-QKD setups without any hardware modifications, as it operates entirely at the software level.


\section*{\label{sec:results}Results}
\begin{figure*}[hbt!]
\center{\includegraphics[scale=0.35]
        {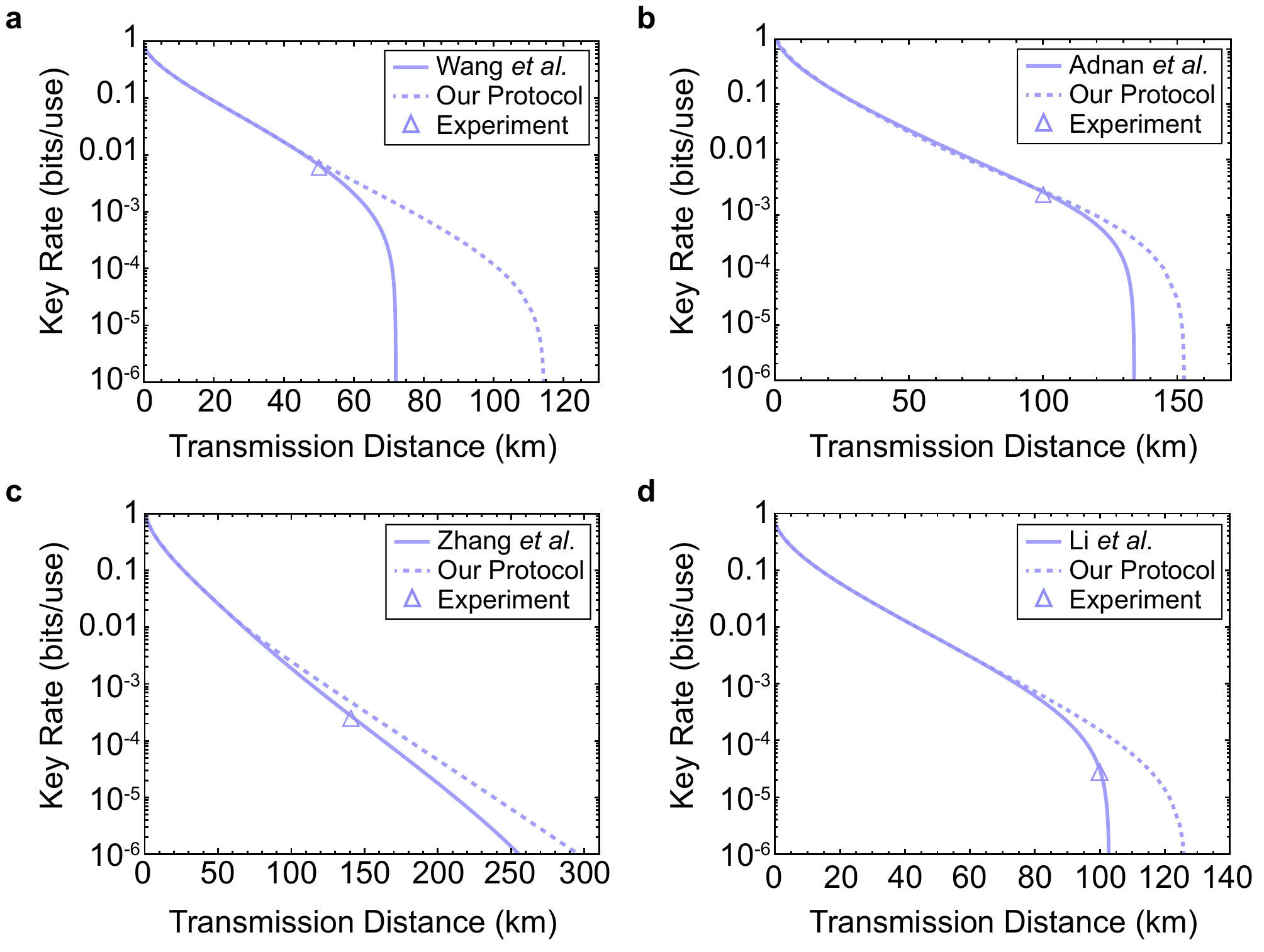}}
\caption{\label{fig:figure5} Simulated key rates for state-of-the-art experiments extended beyond their original demonstration distances. Solid lines indicate the projected key rates using the experimental parameters, while dashed lines show the key rates when applying our protocol with these same parameters. Simulation results are presented for Wang \textit{et al.}~\cite{wang201525} (42 km improvement), Hajomer \textit{et al.}~\cite{hajomer2024long} (19 km improvement), Zhang \textit{et al.}~\cite{zhang2020long} (42 km improvement), and Li \textit{et al.}~\cite{li2023continuous} (23 km improvement) in \textbf{(a)}, \textbf{(b)}, \textbf{(c)}, and \textbf{(d)}, respectively. Refer to Table~\ref{tab:all_experimental_parameters} for the experimental parameters.
}
\end{figure*}
Figure~\ref{fig:figure2} shows the experimental results for a loss of $T=0.26$ with thermal noise values of $W_x=1.02$~SNU and $W_p=1.01$~SNU for the $x$ and $p$ quadratures, respectively. This loss corresponds to an equivalent fibre distance of 29.1 km assuming a fibre loss of 0.2 dB per km. The key rate extracted from the raw data is initially negative, as illustrated in Fig.~\ref{fig:figure2}(a). This is because the modulation variance used in the experiment is not optimal for the $29.1$ km distance with the given noise parameters. The experimental modulation variances for the $x$ and $p$ quadratures were $V_{mod_x} = 12.73$~SNU and $V_{mod_p} = 13.52$~SNU, while the optimal GG02 modulation variances required for this regime are $V_{mod_x} = 6.14$~SNU and $V_{mod_p} = 5.97$~SNU. The asymmetry in the variances is due to asymmetric noise in the state preparation stage. A portion of this preparation noise, primarily due to modulator electronics, is considered trusted in our security analysis (see the Methods). Alice applies a Gaussian filter with gains of $g_x = 0.213$ and $g_p = 0.259$ to reduce the modulation variances. These gains yield the optimal key rate, as indicated in Fig.~\ref{fig:figure2}(b). Although the key rate becomes positive after Alice's post-selection, it does not reach the optimal GG02 key rate as shown in Fig.~\ref{fig:figure2}(a) due to the success probability penalty of the post-selection process ($P_{A_x}=0.68$ and $P_{A_p}=0.60$ are the probability of success of Alice's post-selection on the $x$ and $p$ quadratures, respectively, giving a total probability of success of 0.41). 

\begin{figure*}[hbt!]
\center{\includegraphics[scale=0.4]
        {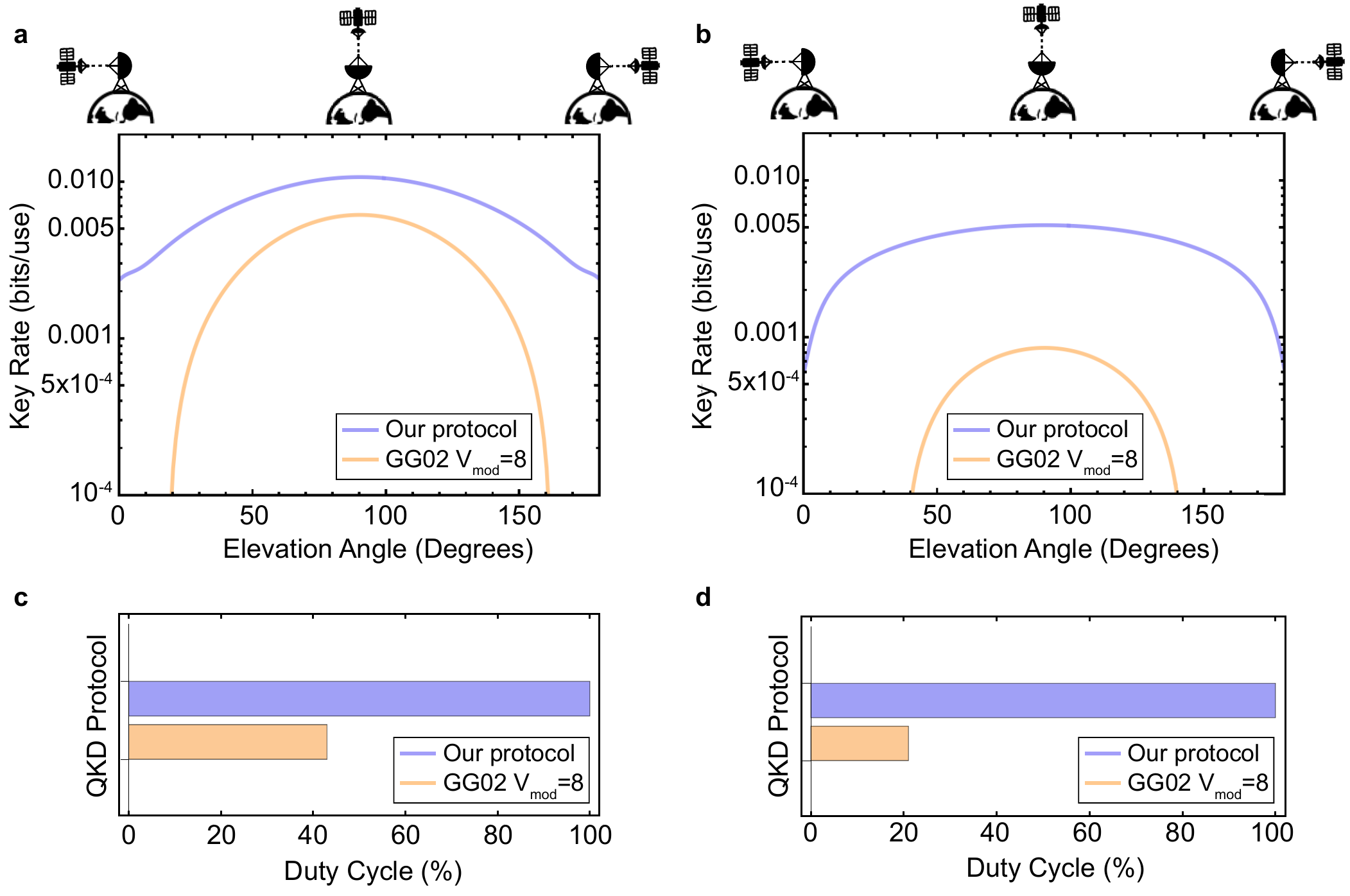}}
\caption{\label{fig:figure6} Key rates for satellite-to-ground communication as a function of elevation angle, using our protocol applied to the GG02 scheme with homodyne detection, assuming the satellite orbits at an altitude of 500 km in low-Earth orbit and the optical ground station is at an elevation of 0 km. We used the model and parameters from Sayat \textit{et al.}~\cite{sayat2024satellite} with a reconciliation efficiency of $\beta=90\%$ (Refer to the \ZJ{Supplementary Information IV} for the parameters). \textbf{(a)} The orange lines represent key rates with a fixed modulation variance of $V_{mod}=8$~SNU, while the purple lines show key rates using our protocol with Alice's post-selection only. This simulation assumes good weather conditions with a visibility of $V=200$ km and low turbulence corresponding to $C_n^2=10^{-16}$ m$^{-2/3}$. At each elevation angle, the \ZJ{post-selection} gain was optimised to achieve the best key rates. \textbf{(b)} Key rates under bad weather conditions with a visibility of $V=20$ km and high turbulence where $C_n^2=10^{-13}$ m$^{-2/3}$.  \textbf{(c)} The duty cycle of our protocol and the GG02 protocol in good weather conditions using a key rate threshold of $10^{-4}$ from (a). \textbf{(d)} The duty cycle of our protocol and the GG02 protocol in bad weather conditions using a key rate threshold of $10^{-4}$ from (b).}
\end{figure*}

To further optimise the key rate, Bob applies a non-Gaussian filter with cut-off parameters of $c_x = 0$ and $c_p = 8.95$ which is shown in Fig.~\ref{fig:figure2}(c). Bob keeps the $x$ quadrature unchanged as it is already optimal but discards most measurements in the $p$ quadrature where the maximum value is $p_b = 9$. The key rate increases and saturates at $0.0038\pm0.001$~bits/use, surpassing the optimal GG02 key rate of $0.0036$~bits/use. 
Our protocol turned regions with negative key rates, deemed insecure by the GG02 protocol, into regions with positive key rates. In these regions, we achieved key rates close to the optimal GG02 rates \ZJ{that are only achievable using optimal modulation which requires accurate channel estimations.}

Next, we investigate distinct regimes where the system benefits from varying post-selection strategies. Figures~\ref{fig:figure3}(a) and (b) show that at 6 km and 15.1 km, 
the key rates with Alice's Gaussian filter match those without post-selection, as the system parameters were already optimal. In this protocol, Alice's Gaussian filter attenuates the quantum signal by reducing the modulation variance, which is useful at long distances where photon losses require smaller optimal variances for the GG02 protocol. For shorter distances, where higher modulation variances are preferred, an inverted Gaussian filter can be used to increase the effective variance and optimise key rates. In contrast to the 6 km and 15.1 km regimes, Alice and Bob's post-selection improves the key rate when experimental parameters are sub-optimal. For example, without post-selection, the chosen parameters yield no positive key rate (solid line, Fig.~\ref{fig:figure3}(c)), but Alice's post-selection makes the key rate positive, surpassing the optimal GG02 rate when Bob also applies post-selection. Similarly, at 39.1 km (transmission of $T=0.17$), Alice's post-selection increased the key rate by 1.7 times (Fig.~\ref{fig:figure3}(d)), but Bob's post-selection did not lead to further improvement, as it is more effective in extreme loss regimes where the GG02 protocol approaches negative key rates.



Furthermore, we applied our protocol to state-of-the-art CV-QKD experimental demonstrations and \ZJ{showed improved performance (see Fig.~\ref{fig:figure4}(a)).} Our protocol increased the key rates of the experimental results of Wang \textit{et al.}~\cite{wang201525} and Hajomer \textit{et al.}~\cite{hajomer2024long}, while doubling the key rate of Zhang \textit{et al.}~\cite{zhang2020long} and achieving a fivefold improvement for Li \textit{et al.}~\cite{li2023continuous} using Alice's post-selection where the percentage improvements are shown in Table~\ref{tab:all_experimental_parameters}. We projected their experimental demonstrations to larger distances by simulating their protocols using the parameters listed in Table~\ref{tab:all_experimental_parameters}, assuming these parameters remain constant across all distances. Under these conditions, our protocol extended the effective transmission distance by 42 km for Wang \textit{et al.}~\cite{wang201525} and Zhang \textit{et al.}~\cite{zhang2020long}, while also enhancing the transmission distance for Hajomer \textit{et al.}~\cite{hajomer2024long} and Li \textit{et al.}~\cite{li2023continuous} by 19 km and 23 km, respectively as shown in Fig.~\ref{fig:figure5}.

Our protocol was also implemented on the Quintessence Labs qOptica system at Data61, CSIRO Sydney, \ZJ{for a fibre distance of 41.71 km}, using heterodyne detection with a modulation variance of approximately 4. Since this variance is close to the optimal value, Alice's post-selection offered little improvement. However, Bob's post-selection tripled the original key rate, surpassing the optimal GG02 rate by emulating a better performing channel as shown in Fig.~\ref{fig:figure4}(b). This improvement can be explained by considering an entanglement-based experiment where Alice creates a two-mode squeezed vacuum state, sending one arm to Bob and measuring the other. With a Gaussian filter, the best achievable key rate is the optimal GG02 rate, as Gaussian operations cannot distill entanglement~\cite{eisert2002distilling}. Non-Gaussian operations, however, enable entanglement distillation, allowing Bob to outperform the GG02 key rate. This was also observed in Hosseinidehaj \textit{et al.}~\cite{hosseinidehaj2020finite} when the MB-NLA filter was made slightly non-Gaussian. However, since their approach emulates a physical filter, it cannot be made highly non-Gaussian without deviating from its intended function as a noiseless amplifier. In contrast, our protocol offers the flexibility to choose any filter cut-off.

Our protocol's applicability extends beyond terrestrial setups and can be applied to \ZJ{improve} satellite-to-ground communication links for CV-QKD. Figure~\ref{fig:figure6}(a) and (b) show how our protocol expands the communication window between the ground station and satellite under varying weather conditions, using the satellite-to-ground link model of Sayat \textit{et al.}~\cite{sayat2024satellite}. Assuming the satellite is in Low Earth orbit at 500 km altitude, the orange lines represent the conventional GG02 protocol with a modulation variance of $V_{mod} = 8$~SNU, while the purple lines show the results of our protocol with Alice’s post-selection. In good weather (Fig.~\ref{fig:figure6}(a)), the GG02 protocol supports communication between the optical ground station (OGS) and satellite at angles from $20^{\circ}$ to $160^{\circ}$. In contrast, Alice’s post-selection enables communication at all angles, boosting key rates by up to 40 times at the extremes where GG02 achieves the lowest key rates. In poor weather (Fig.~\ref{fig:figure6}(b)), our protocol further expands the communication window by $76^{\circ}$, achieving a 400-fold improvement in key rates under extreme conditions. In addition to higher key rates, our protocol offers a longer active communication window, defined as the time during which a positive key rate above $10^{-4}$ bits/use is maintained. Traditional CV-QKD protocols using LEO satellites allow for windows of 1.28 hours in favourable weather and 38 minutes in poor conditions. In contrast, our protocol extends this window to 3 hours in any weather, maintaining a minimum key rate of $10^{-4}$ bits/use. If a higher key rate threshold were used, the GG02 protocol would have even smaller windows, while our protocol would still provide significantly longer active times~(For more details, refer Supplementary Information IV). This showcases that our protocol can be particularly useful in free-space channels, such as satellite-to-ground communication links, where rapid changes in the quantum channel are common.

\section*{\label{sec:discussion}Discussion}
QKD plays a crucial role in securing communications against potential eavesdropping in the quantum era. However, challenges such as limited key rates and vulnerability to environmental factors have impeded its practical deployment.

Addressing these issues, we introduced a new protocol, which enhances CV-QKD by combining post-selection at both Alice and Bob's stations. This protocol utilises a Gaussian filter at Alice's end and a non-Gaussian box-like filter at Bob's end to optimise key rates even under non-ideal conditions. Our approach dynamically adjusts to varying quantum channel conditions, making it particularly effective in free-space channels such as ground-to-satellite communication links.

Experimental results demonstrated significant improvements in key rates when applying our protocol to state-of-the-art CV-QKD systems. We observed up to a five-fold increase in key rates using Alice's post-selection and tripling of the original key rate with Bob's post-selection on different systems. Additionally, our protocol extended the communication window between ground stations and satellites, showing robustness under both good and adverse weather conditions. Specifically, under poor weather conditions, our protocol expanded the communication window to all elevation angles and achieved a $400$-fold improvement in key rates under extreme conditions.

The ability of our protocol to achieve secure key distribution even in rapidly changing quantum channels highlights its potential for practical implementation in challenging environments. By leveraging software-level modifications without the need for hardware changes, our approach provides a flexible and efficient solution for advancing CV-QKD technology.

Our current work focuses on asymptotic key rates, with potential extensions to account for finite-size effects and composable security. In practical applications, data sizes are finite, and smaller block sizes can adversely affect key rates. Nevertheless, our protocol can be effective in scenarios where block sizes degrade key rate quality, however, a full composable security analysis remains an important direction for future work. As Bob's post-selection involves a non-Gaussian filter, the security analysis cannot follow the Gaussian extremality. \textcolor{black}{Instead, Eve's information is bounded by representing her state as a density matrix with some finite truncation in the Fock-number basis.} This is computationally demanding and can be challenging if Alice's variance is too high. A recent paper by Denys \textit{et al.}~\cite{denys2021explicit} provides an analytical lower bound on the asymptotic secret key rate for CV-QKD with arbitrary modulation schemes. Extending this to include Bob's post-selection could bypass the need for density matrices, making the security analysis less time consuming. In this work, we restrict Eve to Gaussian attacks. However, the optimal eavesdropping strategy in the presence of non-Gaussian post-selection remains an open question. Future studies should consider whether Eve could perform a non-Gaussian attack to compromise the key.

Our protocol could potentially be applied to CV-QKD protocols with discrete modulation (DM CV-QKD)~\cite{leverrier2009unconditional, ghorai2019asymptotic, tian2023high}, which utilise a constellation of modulated coherent states. In particular, Bob's non-Gaussian post-selection could enable dynamic switching between different types of DM CV-QKD protocols by selectively combining states from the constellation of modulated coherent states. While Bob's post-selection filter is similar in shape to those used in the DM CV-QKD protocol proposed by Kanitschar and Pacher~\cite{kanitschar2022optimizing}, which demonstrated switching between formats like 8PSK and QPSK, our approach differs in that Alice sends Gaussian-modulated states. This allows Bob to emulate a wider range of modulation formats (e.g., BPSK, QPSK, or multiring) through flexible filter design, enabling more general DM or GM CV-QKD switching.


\section*{\label{sec:methods}Methods}
\noindent \textbf{Alice's Gaussian post-selection.} The random variables generated by Alice's function generator follow a Gaussian distribution as described below
\begin{equation}
    \label{eq:alices_dist}
    p_a(x_a)=\frac{1}{\sqrt{2\pi V_{mod_x}}}\mathrm{exp}\bigg[\frac{-x_a^2}{2V_{mod_x}}\bigg],
\end{equation}
where $V_{mod_x}$ represents the modulation variance of Alice along the $x$ quadrature with random variables $x_a$. The modulation variance of the $x$ and $p$ quadrature are not symmetric due to the experimental imperfections. However, as Bob performs a homodyne measurement, Alice's Gaussian distribution of the $x$ and $p$ quadrature variables can be expressed individually. \ZJ{In the following, we explicitly present the analysis for the $x$ quadrature, which can be similarly applied to the $p$ quadrature.} When Alice applies the filter shown in Eq.~\eqref{eq:alices_filter}, the probability of success is given by
\begin{equation}
    \label{eq:alices_ps}
    P_{A_x}=\ZJ{\int_{-\infty}^\infty F_A(x_a)p_a(x_a)\,d{x_a}}=\frac{1}{\sqrt{2g_x^2V_{mod_x}+1}},
\end{equation}
where $g_x$ is the gain of the Gaussian filter for the $x$ quadrature. \textcolor{black}{Experimentally, Alice's filter, as described in Eq.~\eqref{eq:alices_filter}, is implemented by generating a random number uniformly distributed between 0 and 1. The filter then retains the variables whose corresponding filter function values are greater than or equal to this random number.} The probability of success is determined by the ratio of samples kept after filtering to the total number of samples. For our experiment, the gains were optimised to maximise the key rates. For a distance of 29.1 km, we used $g_x = 0.213$ and $g_p = 0.259$, while for a distance of 39.1 km, the gains were $g_x = 0.178$ and $g_p = 0.213$.

After Alice's post-selection, the modulation variance of the effective thermal state is reduced. The equivalent modulation variance is expressed as follows
\begin{equation}
    \label{eq:new_modulation_variance}
    \Tilde{V}_{mod_x}=\ZJ{\int_{-\infty}^\infty \!\!\frac{x_a^2F_A(x_a)p_a(x_a)}{P_{A_x}}\,d{x_a}}=\frac{V_{mod_x}}{2g_x^2 V_{mod_x}+1}.
\end{equation}
Alice's post-selection cannot modify the channel parameters. However, Bob's effective variance also reduces, as he must disregard the data that Alice discards. Therefore, \ZJ{Bob's variance, using Alice's new variance $V_x=\Tilde{V}_{mod_x}+1$, can be expressed as}
\begin{equation}
    \label{eq:bobs_variance}
    V_{b_x}=T(V_x+\xi_x)+(1-T)W_x,
\end{equation}
where $\xi_x$ is trusted excess noise coming from the state preparation, $T$ is the channel transmittance that \ozlem{is} introduced experimentally using a half-wave plate followed by a PBS. This loss is equivalent to a distance of $d=-\log_{10}T/0.02$ assuming a loss of $0.2$ dB per km. $W_x$ denotes the thermal noise from the quantum channel, attributed to Eve. Similarly, the covariance between Alice and Bob after Alice's post-selection scales to $T\Tilde{V}_{mod_x}$ in the prepare-and-measure (PM) scheme. QKD protocols can be expressed in either entanglement-based (EB) or prepare-and-measure schemes. While mathematically equivalent~\cite{weedbrook2012gaussian,grosshans2003virtual}, the \ZJ{EB} representation is more convenient for security analysis. Therefore, the new covariance between Alice and Bob in the equivalent EB scheme can be expressed as 
\begin{equation}
    \label{eq:new_covariance}
    C_x=\sqrt{T(V_x^2-1)}.
\end{equation}

The EB-based covariance matrix of the quantum state between Alice and Bob can be written as~\cite{laudenbach2018continuous} \ZJ{(Refer to Supplementary Information I for the derivation of the PM covariance matrix and Supplementary Information III for derivation of the covariance matrices starting from the EB model})
\begin{equation}
    \label{eq:new_CM_after_alice}
   \ZJ{\sigma_{AB}^{\text{EB}}}=\begin{pmatrix}
    V_x & 0 & C_x & 0 \\
    0 & V_p & 0 & -C_p \\
    C_x & 0 & V_{b_x} & 0 \\
    0 & -C_p & 0 & V_{b_p}
    \end{pmatrix}.
\end{equation}
Note that $C_p$ has the same form as $C_x$, but with a different variance, $V_p$. Similarly, Bob's variance $V_{b_p}$ follows the same formula as Eq.~\eqref{eq:bobs_variance}, with the variance $V_p$, trusted noise $\xi_p$, and thermal noise $W_p$.

The mutual information between Alice and Bob can be calculated using the EB-based covariance matrix \ZJ{shown in Eq.~\eqref{eq:new_CM_after_alice}}. In our protocol, we employ reverse reconciliation, meaning the direction of classical communication is from Bob to Alice~\cite{garcia2007quantum}. While the direction of reconciliation does not affect the mutual information between Alice and Bob, it influences how much information Eve can access. CV-QKD protocols with reverse reconciliation, unlike direct reconciliation, allow key distribution beyond the 3 dB loss limit~\cite{grosshans2002reverse, grosshans2003quantum}. The mutual information between Alice and Bob is calculated from
\begin{equation}
    \label{eq:mutual_info_after_alice}
    I_{AB_x}=\frac{1}{2}\log_2\frac{V_{A_x}}{V_{A_x|B_x}},
\end{equation}
\ZJ{where $V_{A_x}$ represents Alice's variance of the classical data in the EB scheme where she performs a heterodyne measurement on the mode she keeps. This variance is given by $V_{A_x} = (V_x + 1)/2$ and $V_{A_x|B_x} = V_{A_x} - C_x^2/(2V_{b_x})$ denotes Alice’s variance conditioned on Bob’s measurement outcomes.}

Eve's average covariance matrix also scales after Alice's post-selection, as Alice and Bob keep fewer states than originally. As previously mentioned, 
Eve injects a TMSV state with modes $E_1$ and $E_2$. Mode $E_1$ interacts with incoming signal from Alice, while mode $E_2$ is retained by Eve. After Alice's post-selection, the variance of $E_1$ is updated as follows, while the variance of $E_2$ remains unchanged
\begin{equation}
    \label{eq:eves_variance}
    V_{E_{1_x}}=TW_x+(1-T)(V_x+\xi_x).
\end{equation}
Note that the variance of the $p$ quadrature for this mode follows the same formula, but with Alice's variance $V_p$, trusted excess noise $\xi_p$ and a thermal noise of $W_p$. \ZJ{In practice, there exists a slight discrepancy between $W_{x}$ and $W_{p}$} due to the asymmetry in the experimental measurements. This implies that Eve introduces slightly different noise in the $x$ and $p$ quadratures, expressed as $W_x$ and $W_p$, respectively. To achieve this, Eve must create two separate TMSV states~\cite{weedbrook2012gaussian} and combine them at a 50/50 beamsplitter before interacting with Alice's signal. Based on these, Eve's average covariance matrix can be expressed as
\begin{equation}
    \label{eq:eves_average_CM}
    \sigma_{E_{1}E_{2}}=\begin{pmatrix}
    V_{E_{1_x}} & 0 & C_{E_x} & 0 \\
    0 & V_{E_{1_p}} & 0 & C_{E_p} \\
    C_{E_x} & 0 & W_x & 0 \\
    0 & C_{E_p} & 0 & W_p
    \end{pmatrix},
\end{equation}
where $C_{E_x}=0.5(-\mathrm{e}^{2r_1}+\mathrm{e}^{2r_2})$ and $C_{E_p}=0.5(-\mathrm{e}^{-2r_1}+\mathrm{e}^{-2r_2})$ where $r_1$ and $r_2$ are the squeezing parameters of the TMSV states of Eve which are given by $r_1=0.5\ln({W_x+\sqrt{W_x^2-W_x/W_p}})$ and $r_2=0.5\ln({W_x/W_p})-0.5\ln({W_x+\sqrt{W_x^2-W_x/W_p}})$. Eve's information is calculated using the Holevo bound~\cite{holevo1998capacity}, which is used to upper bound the information leaked to Eve as shown below
\begin{equation}
    \label{eq:holevo_bound}
    I_{E_x}=S(\sigma_{E_1E_2})-S(\sigma_{E_1E_2|B_x}),
\end{equation}
where $S(\sigma_{E_1E_2})$ and $S(\sigma_{E_1E_2|B_x})$ represent the von Neumann entropy of Eve's average state and Eve's state conditioned on Bob's measurement outcome $x_b$, \ZJ{respectively}. Eve's conditional \ZJ{covariance matrix} can be found from~\cite{laudenbach2018continuous}
\begin{equation}
    \label{eq:eves_conditional_state}
    \sigma_{E_1E_2|B_x}=\sigma_{E_1E_2}-\frac{1}{V_{b_x}}\sigma_c\Pi_x\sigma_c^T,
\end{equation}
where $\Pi_x=\mathrm{diag}(1,0)$ if \ZJ{Bob measures $x$ quadrature} and $\Pi_p=\mathrm{diag}(0,1)$ for $p$ quadrature. \ZJ{$\sigma_c$ is the covariance terms between Eve and Bob}, given as
\begin{equation}
    \sigma_c=\begin{pmatrix}
        C_{E_1B_x} & 0 \\
        0 & C_{E_1B_p} \\
        C_{E_2B_x} & 0 \\
        0 & C_{E_2B_p}
    \end{pmatrix},
\end{equation}
where $C_{E_1B_x}=\sqrt{T(1-T)}\big(W_x-(V_x+\xi_x)\big)$, $C_{E_1B_p}=\sqrt{T(1-T)}\big(W_p-(V_p+\xi_p)\big)$, $C_{E_2B_x}=\sqrt{(1-T)}C_{E_x}$ and $C_{E_2B_p}=\sqrt{(1-T)}C_{E_p}$. Refer to Laudenbach \textit{et al.}~\cite{laudenbach2018continuous} for the full derivation of $\sigma_c$.

The secret key rate is calculated using Eqs.~\eqref{eq:mutual_info_after_alice} and ~\eqref{eq:holevo_bound}:
\begin{equation}
    \label{eq:keyrate_equation}
    K=\beta I_{AB}-I_E,
\end{equation}
where $\beta$ represents the reconciliation efficiency, with $0 \leq \beta < 1$. \ZJ{In our experiment, we assume $\beta = 92\%$ when calculating the key rates.}

As Bob performs homodyne measurements on the $x$ quadrature half of the time and on the $p$ quadrature the other half, and since the variances and noise parameters differ between these quadratures due to experimental imperfections, Alice and Bob's mutual information, as well as Eve's information, must be averaged over both quadratures. This leads to $I_{AB}=0.5 P_{A_x}\times I_{AB_x}+0.5 P_{A_p}\times I_{AB_p}$ and $I_E=0.5 P_{A_x}\times I_{E_x}+0.5 P_{A_p}\times I_{E_p}$.

\noindent \textbf{Bob's non-Gaussian post-selection.} Bob's initial distribution before his post-selection can be expressed as 
\begin{equation}
    \label{eq:bobs_distribution}
    p_b(x_b)=\frac{1}{\sqrt{2\pi V_{b_x}}}\mathrm{exp}\bigg[\frac{-x_b^2}{2V_{b_x}}\bigg],
\end{equation}
where $x_b$ corresponds to Bob's measurement outcome. Following Alice's post-selection, Bob applies the filter shown in Eq.~\eqref{eq:bobs_filter}, which gives a probability of success of
\begin{equation}
    \label{eq:bobs_ps}
    P_{B_x}=\ZJ{\int_{-\infty}^\infty F_B(x_b)p_b(x_b)\,d{x_b}}=1-\mathrm{erf}\bigg[\frac{c_x}{\sqrt{2V_{b_x}}}\bigg],
\end{equation}
\ZJ{where $c_x$ is the cut-off parameter of the filter function.}

Following Bob's \ZJ{post-selection}, his output probability distribution \ZJ{becomes}
\begin{equation}
    \Tilde{p}_b(x_b)=\begin{cases}
        0 & \text{$-c_x<x_{b}<c_x$} \\
        \frac{p_b(x_b)}{P_{B_x}} & \text{$\mathrm{elsewhere}$}.
    \end{cases}
    \label{eq:bobs_output_distribuion}
\end{equation}
As Bob's filter is non-Gaussian, we avoid using the Gaussian extremality as this approach underestimates the secret key rate due to overestimating Eve’s information. Therefore, we propose calculating the absolute bound on Eve’s information directly from her density matrix.

To express Eve's conditional state shown in Eq.~\eqref{eq:eves_conditional_state} as a density matrix, we first use Williamson's decomposition, which provides a symplectic matrix $s$ and a diagonal matrix $\omega$, following $s\sigma_{E_1E_2|B_x} s^T = \omega$~\cite{williamson1936algebraic}. We then decompose the symplectic matrix, $s$, further using the Bloch-Messiah decomposition~\cite{cariolaro2016reexamination}, representing $s$ as a series of beamsplitters, squeezers, and phase rotations: $s=s_us_is_v$ in the covariance matrix notation and $S=S_US_IS_V$ in the density matrix notation. In our case, both $s_u$ and $s_v$ are expressed through phase rotations and beamsplitters: $S_U=R(\theta_1, \theta_2) B(\tau_U) R(\theta_3, \theta_4)$ and $S_V=R(\theta_5, \theta_6) B(\tau_V) R(\theta_7, \theta_8)$, where each parameter needs to be solved. The $s_i$ component is expressed through \ZJ{two-mode} squeezing operations, $S_I=S(r_1, r_2)$. Therefore, Eve's overall conditional density matrix is written in terms of the symplectic density matrix $S$ and a mixed thermal state $\Lambda$
\begin{equation}
    \label{eq:eves_density_matrix}
    \rho_{E_{1}E_{2}|B_x}(x_b=0)=S \Lambda S^\dagger,
\end{equation}
where $\Lambda = \rho_1 \otimes \rho_2$, with $\rho_1$ and $\rho_2$ being thermal states given by $\sum\limits_{n=0}^{\infty} \frac{\bar{n}^n}{(\bar{n}+1)^{n+1}} \ket{n}\!\bra{n}$, having mean photon numbers $\bar{n}_1$ and $\bar{n}_2$, respectively. Note that \ZJ{$\bar{n}_i = (\lambda_i - 1)/2$}, where \ZJ{$\lambda_i$} represents \ZJ{the $i^\text{th}$} diagonal terms of the diagonal matrix $\omega$. In our case, $\omega$ has two unique diagonal terms, $\lambda_1$ and $\lambda_2$, which are used to calculate the mean photon numbers $\bar{n}_1$ and $\bar{n}_2$.

To calculate Eve's information, we need to determine her average density matrix, which is the sum of her all conditional matrices \ZJ{over} Bob's measurement outcomes \ZJ{weighted by the probability density for the respective outcome}. \ZJ{Given that Bob measures} an outcome $x_b$, Eve's conditional state is
\begin{multline}
\label{eq:displaced_conditional_state}
        \rho_{E_{1}E_{2}|B_x}(x_b)=D(\alpha_1({x_{b})},\alpha_2({x_{b}}))\rho_{E_{1}E_{2}|B_x}(x_b=0)\\D^\dagger(\alpha_1({x_{b})},\alpha_2({x_{b}})),
\end{multline}
where $D(\alpha_1({x_{b})},\alpha_2({x_{b}}))$ represents the displacement operator for Eve's modes 1 and 2 in the form of 
\begin{align}
    \label{eq:displacement_operator}
    D(\alpha_1({x_{b})},\alpha_2({x_{b}}))=&\exp(\alpha_1(x_b)a^\dagger-\alpha_1(x_b)a)\otimes\nonumber \\&\exp(\alpha_2(x_b)a^\dagger-\alpha_2(x_b)a),
\end{align}
where $\alpha_1$ and $\alpha_2$ are functions of Eve's conditional means based on Bob's measurement outcomes, $x_b$. \ZJ{The matrix of Eve's conditional means is calculated from}
\begin{equation}
\label{eq:conditional_means_homodyne}
    \mu_{E_1E_2}(x_b,p_b)=\sigma_c\begin{pmatrix}
        V_{b_x} & 0\\
        0 & V_{b_p}
    \end{pmatrix}^{-1}
    \begin{pmatrix}
       x_b \\ p_b
    \end{pmatrix},
\end{equation}
where $p_b=0$ for homodyne detection when $x_b$ is measured and $x_b=0$ when $p_b$ is measured. Therefore $\alpha_1(x_b)=\mu_{E_1}(x_b)/2$ and $\alpha_2(x_b)=\mu_{E_2}(x_b)/2$ where \ZJ{$\mu_{E_1}(x_b)$ and $\mu_{E_2}(x_b)$ are obtained from the entries of the $\mu_{E_1E_2}(x_b,0)$ vector as follows:} $\mu_{E_1}(x_b)=\mu_{E_1E_2}(x_b,0)_1$ and $\mu_{E_2}(x_b)=\mu_{E_1E_2}(x_b,0)_3$.

Although each conditional state has the same entropy, they contribute differently to Eve's average state due to the varying probabilities of obtaining each conditional state. Therefore, we determine Eve's average density matrix by multiplying each conditional state with their corresponding probabilities from Bob's post-selected probability distribution followed by summing them all up. Note that Bob's probability of getting a given outcome is obtained from 
\begin{equation}
    p_b'(x_i)=\int_{x_i-\Delta}^{x_i+\Delta} \Tilde{p}_b(x_i) \, dx_b.
    \label{eq:bobs_probability_pxi2}
\end{equation}
To calculate the key rates after the non-Gaussian post-selection, we discretise Eve's information to determine the conditional states for each of Bob's measurement outcomes. \textcolor{black}{Therefore, $\Delta$ in Eq.~\eqref{eq:bobs_probability_pxi2} is the width of the discretised bins and in our results $\Delta\!=\!0.1$.} Consequently, Eve's average state can be calculated from Eq.~\eqref{eq:eve_ave_denmat}. Eve's information is then calculated using the Holevo bound which is
\begin{equation}
    I_{E_x}=S(\rho_{E_1E_2})-S(\rho_{E_{1}E_{2}|B_x}(0)),
    \label{eq:eves_information_density_matrix}
\end{equation}
where $S(\rho_{E_1E_2})$ represents the von Neumann entropy for Eve's average state while $S(\rho_{E_{1}E_{2}|B_x}(0))$ gives the von Neumann entropy of Eve's conditional state \ZJ{which remains the same irrespective of Bob's measurement outcome $x_b$.}

In order to find Alice and Bob's mutual information, the new bivariate Gaussian distribution between Alice and Bob following Alice's post-selection is fitted to the following function
\begin{multline}
    f_{AB}(x_a,x_b)=\frac{1}{2\pi\sigma_a\sigma_b\sqrt{1-\rho^2}}\\\exp\bigg[-\frac{1}{2(1-\rho^2)}\bigg(\frac{x_a^2}{\sigma_a^2}-\frac{2\rho x_ax_b}{\sigma_a\sigma_b}+\frac{x_b^2}{\sigma_b^2}\bigg)\bigg],
    \label{eq:multivariate_distribution}
\end{multline}
where $\sigma_a = \sqrt{(V_x + 1)/2}$ and $\sigma_b = \sqrt{V_{b_x}}$ are the standard deviations of Alice and Bob, respectively, and $\rho = C_x / \sqrt{2\sigma_a\sigma_b}$ represents the correlation term between them. 

Using Eq.~\eqref{eq:multivariate_distribution}, we generate a probability table between Alice and Bob for each values of $x_a$ and $x_b$ where $x_a=-15,-15+\Delta,\cdots,15$ and $x_b=-9,-9+\Delta,\cdots,9$. The mutual information then is calculated from
\begin{equation}
    I_{AB_x}=H_{A_x}+H_{B_x}-H_{AB_x},
    \label{eq:iab_after_bob}
\end{equation}
where $H_{A_x}$, $H_{B_x}$, and $H_{AB_x}$ represent the Shannon entropy of Alice's classical variable, Bob's classical variable, and the joint entropy of Alice's and Bob's classical variables, respectively.

After Bob's post-selection $H_{AB_x}$ can be computed from
\begin{multline}
    H_{AB_x}=-\sum_{x_a=-n_a}^{n_a}\sum_{x_b=-n_b}^{-c_x-\Delta/2}\frac{f_{AB}(x_a,x_b)}{P_{B_x}}\\\log_2\bigg[\frac{f_{AB}(x_a,x_b)}{P_{B_x}}\bigg]-\sum_{x_a=-n_a}^{n_a}\sum_{x_b=c_x+\Delta/2}^{n_b}\frac{f_{AB}(x_a,x_b)}{P_{B_x}}\\\log_2\bigg[\frac{f_{AB}(x_a,x_b)}{P_{B_x}}\bigg],
    \label{eq:joint_entropy}
\end{multline}
\ZJ{where $c_x$ is the cut-off parameter of Bob's filter function in Eq.~\eqref{eq:bobs_filter}}; where $n_a$ and $n_b$ represent the maximum number that Alice prepares and Bob measures, respectively.

Alice's probabilities after Bob's post-selection can be found from
\begin{equation}
    f_A(x_a)=\sum_{x_b=-n_b}^{-c_x-\Delta/2}\frac{f_{AB}(x_a,x_b)}{P_{B_x}}+\sum_{x_b=c_x+\Delta/2}^{n_b}\frac{f_{AB}(x_a,x_b)}{P_{B_x}},
    \label{eq:alices_prob_after_bob}
\end{equation}
which is used to compute $H_{A_x}$
\begin{equation}
    H_{A_x}=-\sum_{x_a=-n_a}^{n_a}f_A(x_a)\log_2\big[f_A(x_a)\big].
    \label{eq:alices_entropy_after_bob}
\end{equation}

Similarly, Bob's probabilities from the probability table after his post-selection can be found from
\begin{equation}
    f_B(x_b)=\sum_{x_a=-n_a}^{n_a}\frac{f_{AB}(x_a,x_b)}{P_{B_x}},
    \label{eq:bobs_prob_after_bob}
\end{equation}
which is used to calculate $H_{B_x}$ as
\begin{multline}
    H_{B_x}=-\sum_{x_b=-n_b}^{-c_x-\Delta/2}f_B(x_b)\log_2\big[f_B(x_b)\big]-\\\sum_{x_b=c_x+\Delta/2}^{n_b}f_B(x_b)\log_2\big[f_B(x_b)\big].
    \label{eq:bobs_entropy_after_bob}
\end{multline}

\noindent \textbf{Calculating the final key rate.} Since both Alice and Bob perform post-selection, we need to consider the probabilities of success and average the key rates over the $x$ and $p$ quadratures. Following Bob's filtering process, the average mutual information between Alice and Bob becomes $I_{AB} = 0.5 P_{B_x} P_{A_x} I_{AB_x} + 0.5 P_{B_p} P_{A_p} I_{AB_p}$. Eve's information is averaged in the same manner as the mutual information. Finally, the key rate is calculated using Eq.~\eqref{eq:keyrate_equation}.

\noindent \textbf{Error bars.}
\textcolor{black}{In QKD, key rate calculations typically rely on a worst-case estimation of the channel parameters derived from parameter estimation. In our approach, we fit the raw data to a bivariate Gaussian distribution with the same formula as Eq.~\eqref{eq:multivariate_distribution} within a $95\%$ confidence interval to obtain the values of $\sigma_a$, $\sigma_b$ and $\rho$. In this case, $\sigma_a$ and $\sigma_b$ the standard deviations of Alice's modulation, $\sqrt{V_{mod_x}}$, and Bob's measurement outcomes, $\sqrt{V_{b_x}}$, respectively, while $\rho$ denotes the correlation parameter between Alice and Bob's data.}

\textcolor{black}{Since our data sample is finite, the error bars in the key rates represent statistical errors, calculated by bootstrapping the data. We resample the data 100 times, with 10 million samples in each iteration, effectively simulating the experiment 100 times. The variance across these simulations provides the error bars.}

\noindent \textbf{Data availability:}
The data that supports the findings of this study is available from the corresponding author upon reasonable request.

\noindent \textbf{Code availability:}
The codes that support the findings of this study are available from the corresponding author upon reasonable request.

\noindent \textbf{Acknowledgments:}
We extend our gratitude to Mikhael Sayat for his valuable discussions on satellite communication link budget calculations. \ozlem{We also extend our thanks to Dr.~Aaron Tranter for his support in troubleshooting the opto-electronics involved in this experiment and to Dr.~Oliver Thearle for constructing the detectors, helping with the experimental setup, and assisting with troubleshooting. Lastly, we thank Dr.~Hao Jeng for his valuable contributions to discussions on post-selection and security analysis.} This research was funded by the Australian Research Council Centre of Excellence for Quantum Computation and Communication Technology (Grant No. CE170100012) and by A$^*$STAR grants C230917010~(Emerging Technology), C230917004~(Quantum Sensing) and Q.InC Strategic Research and Translational Thrust.

\noindent \textbf{Author contributions:}
PKL and SMA conceived the project. OE, BS, and SMA developed the protocol. OE, LOC, BS, and JZ constructed the experiment. AW assisted with troubleshooting. OE conducted the experiment and analysed the data. TS set up the experiment at Data61, and SK collected the GG02 data with heterodyne detection. AD provided theoretical help in mapping Eve's covariance matrix to the density matrix. PKL, SMA, and JZ supervised the project. OE drafted the manuscript with contributions from all authors.

\noindent \textbf{Competing interests:}
The authors declare no competing financial or non-financial interests.

\FloatBarrier

\bibliography{apssamp}

\providecommand{\noopsort}[1]{}\providecommand{\singleletter}[1]{#1}%
\begin{thebibliography}{10}
\expandafter\ifx\csname url\endcsname\relax
  \def\url#1{\texttt{#1}}\fi
\expandafter\ifx\csname urlprefix\endcsname\relax\def\urlprefix{URL }\fi
\providecommand{\bibinfo}[2]{#2}
\providecommand{\eprint}[2][]{\url{#2}}

\bibitem{ekert2014}
\bibinfo{author}{Ekert, A.} \& \bibinfo{author}{Renner, R.}
\newblock \bibinfo{title}{The ultimate physical limits of privacy}.
\newblock \emph{\bibinfo{journal}{Nature}} \textbf{\bibinfo{volume}{507}}, \bibinfo{pages}{443--447} (\bibinfo{year}{2014}).

\bibitem{gisin2002}
\bibinfo{author}{Gisin, N.}, \bibinfo{author}{Ribordy, G.}, \bibinfo{author}{Tittel, W.} \& \bibinfo{author}{Zbinden, H.}
\newblock \bibinfo{title}{Quantum cryptography}.
\newblock \emph{\bibinfo{journal}{Rev. Mod. Phys.}} \textbf{\bibinfo{volume}{74}}, \bibinfo{pages}{145--195} (\bibinfo{year}{2002}).

\bibitem{pirandola2020advances}
\bibinfo{author}{Pirandola, S.} \emph{et~al.}
\newblock \bibinfo{title}{Advances in quantum cryptography}.
\newblock \emph{\bibinfo{journal}{Adv. Opt. Photonics}} \textbf{\bibinfo{volume}{12}}, \bibinfo{pages}{1012--1236} (\bibinfo{year}{2020}).

\bibitem{ralph1999}
\bibinfo{author}{Ralph, T.~C.}
\newblock \bibinfo{title}{Continuous variable quantum cryptography}.
\newblock \emph{\bibinfo{journal}{Phys. Rev. A}} \textbf{\bibinfo{volume}{61}}, \bibinfo{pages}{010303} (\bibinfo{year}{1999}).

\bibitem{hillery2000}
\bibinfo{author}{Hillery, M.}
\newblock \bibinfo{title}{Quantum cryptography with squeezed states}.
\newblock \emph{\bibinfo{journal}{Phys. Rev. A}} \textbf{\bibinfo{volume}{61}}, \bibinfo{pages}{022309} (\bibinfo{year}{2000}).

\bibitem{grosshans2002continuous}
\bibinfo{author}{Grosshans, F.} \& \bibinfo{author}{Grangier, P.}
\newblock \bibinfo{title}{Continuous variable quantum cryptography using coherent states}.
\newblock \emph{\bibinfo{journal}{Phys. Rev. Lett.}} \textbf{\bibinfo{volume}{88}}, \bibinfo{pages}{057902} (\bibinfo{year}{2002}).

\bibitem{grosshans2003quantum}
\bibinfo{author}{Grosshans, F.} \emph{et~al.}
\newblock \bibinfo{title}{Quantum key distribution using gaussian-modulated coherent states}.
\newblock \emph{\bibinfo{journal}{Nature}} \textbf{\bibinfo{volume}{421}}, \bibinfo{pages}{238--241} (\bibinfo{year}{2003}).

\bibitem{weedbrook2004quantum}
\bibinfo{author}{Weedbrook, C.} \emph{et~al.}
\newblock \bibinfo{title}{Quantum cryptography without switching}.
\newblock \emph{\bibinfo{journal}{Phys. Rev. Lett.}} \textbf{\bibinfo{volume}{93}}, \bibinfo{pages}{170504} (\bibinfo{year}{2004}).

\bibitem{lance2005no}
\bibinfo{author}{Lance, A.~M.} \emph{et~al.}
\newblock \bibinfo{title}{No-switching quantum key distribution using broadband modulated coherent light}.
\newblock \emph{\bibinfo{journal}{Phys. Rev. Lett.}} \textbf{\bibinfo{volume}{95}}, \bibinfo{pages}{180503} (\bibinfo{year}{2005}).

\bibitem{lodewyck2007quantum}
\bibinfo{author}{Lodewyck, J.} \emph{et~al.}
\newblock \bibinfo{title}{Quantum key distribution over 25 km with an all-fiber continuous-variable system}.
\newblock \emph{\bibinfo{journal}{Phys. Rev. A}} \textbf{\bibinfo{volume}{76}}, \bibinfo{pages}{042305} (\bibinfo{year}{2007}).

\bibitem{madsen2012continuous}
\bibinfo{author}{Madsen, L.~S.}, \bibinfo{author}{Usenko, V.~C.}, \bibinfo{author}{Lassen, M.}, \bibinfo{author}{Filip, R.} \& \bibinfo{author}{Andersen, U.~L.}
\newblock \bibinfo{title}{Continuous variable quantum key distribution with modulated entangled states}.
\newblock \emph{\bibinfo{journal}{Nat. Commun.}} \textbf{\bibinfo{volume}{3}}, \bibinfo{pages}{1083} (\bibinfo{year}{2012}).

\bibitem{jouguet2013experimental}
\bibinfo{author}{Jouguet, P.}, \bibinfo{author}{Kunz-Jacques, S.}, \bibinfo{author}{Leverrier, A.}, \bibinfo{author}{Grangier, P.} \& \bibinfo{author}{Diamanti, E.}
\newblock \bibinfo{title}{Experimental demonstration of long-distance continuous-variable quantum key distribution}.
\newblock \emph{\bibinfo{journal}{Nat. Photonics}} \textbf{\bibinfo{volume}{7}}, \bibinfo{pages}{378--381} (\bibinfo{year}{2013}).

\bibitem{wang201525}
\bibinfo{author}{Wang, C.} \emph{et~al.}
\newblock \bibinfo{title}{25 mhz clock continuous-variable quantum key distribution system over 50 km fiber channel}.
\newblock \emph{\bibinfo{journal}{Sci. Rep.}} \textbf{\bibinfo{volume}{5}}, \bibinfo{pages}{14607} (\bibinfo{year}{2015}).

\bibitem{zhang2020long}
\bibinfo{author}{Zhang, Y.} \emph{et~al.}
\newblock \bibinfo{title}{Long-distance continuous-variable quantum key distribution over 202.81 km of fiber}.
\newblock \emph{\bibinfo{journal}{Phys. Rev. Lett.}} \textbf{\bibinfo{volume}{125}}, \bibinfo{pages}{010502} (\bibinfo{year}{2020}).

\bibitem{jain2022practical}
\bibinfo{author}{Jain, N.} \emph{et~al.}
\newblock \bibinfo{title}{Practical continuous-variable quantum key distribution with composable security}.
\newblock \emph{\bibinfo{journal}{Nat. Commun.}} \textbf{\bibinfo{volume}{13}}, \bibinfo{pages}{4740} (\bibinfo{year}{2022}).

\bibitem{hajomer2024long}
\bibinfo{author}{Hajomer, A.~A.} \emph{et~al.}
\newblock \bibinfo{title}{Long-distance continuous-variable quantum key distribution over 100-km fiber with local local oscillator}.
\newblock \emph{\bibinfo{journal}{Sci. Adv.}} \textbf{\bibinfo{volume}{10}}, \bibinfo{pages}{eadi9474} (\bibinfo{year}{2024}).

\bibitem{garcia2007quantum}
\bibinfo{author}{Garcia-Patron~Sanchez, R.}
\newblock \bibinfo{title}{Quantum information with optical continuous variables: from bell tests to key distribution}  (\bibinfo{year}{2007}).

\bibitem{xiang2010heralded}
\bibinfo{author}{Xiang, G.-Y.}, \bibinfo{author}{Ralph, T.~C.}, \bibinfo{author}{Lund, A.~P.}, \bibinfo{author}{Walk, N.} \& \bibinfo{author}{Pryde, G.~J.}
\newblock \bibinfo{title}{Heralded noiseless linear amplification and distillation of entanglement}.
\newblock \emph{\bibinfo{journal}{Nat. Photonics}} \textbf{\bibinfo{volume}{4}}, \bibinfo{pages}{316--319} (\bibinfo{year}{2010}).

\bibitem{winnel2020generalized}
\bibinfo{author}{Winnel, M.~S.}, \bibinfo{author}{Hosseinidehaj, N.} \& \bibinfo{author}{Ralph, T.~C.}
\newblock \bibinfo{title}{Generalized quantum scissors for noiseless linear amplification}.
\newblock \emph{\bibinfo{journal}{Phys. Rev. A}} \textbf{\bibinfo{volume}{102}}, \bibinfo{pages}{063715} (\bibinfo{year}{2020}).

\bibitem{bencheikh2001quantum}
\bibinfo{author}{Bencheikh, K.}, \bibinfo{author}{Symul, T.}, \bibinfo{author}{Jankovic, A.} \& \bibinfo{author}{Levenson, J.-A.}
\newblock \bibinfo{title}{Quantum key distribution with continuous variables}.
\newblock \emph{\bibinfo{journal}{J. Mod. Opt.}} \textbf{\bibinfo{volume}{48}}, \bibinfo{pages}{1903--1920} (\bibinfo{year}{2001}).

\bibitem{huang2013performance}
\bibinfo{author}{Huang, P.}, \bibinfo{author}{He, G.}, \bibinfo{author}{Fang, J.} \& \bibinfo{author}{Zeng, G.}
\newblock \bibinfo{title}{Performance improvement of continuous-variable quantum key distribution via photon subtraction}.
\newblock \emph{\bibinfo{journal}{Phys. Rev. A}} \textbf{\bibinfo{volume}{87}}, \bibinfo{pages}{012317} (\bibinfo{year}{2013}).

\bibitem{hu2020continuous}
\bibinfo{author}{Hu, L.}, \bibinfo{author}{Al-Amri, M.}, \bibinfo{author}{Liao, Z.} \& \bibinfo{author}{Zubairy, M.}
\newblock \bibinfo{title}{Continuous-variable quantum key distribution with non-gaussian operations}.
\newblock \emph{\bibinfo{journal}{Phys. Rev. A}} \textbf{\bibinfo{volume}{102}}, \bibinfo{pages}{012608} (\bibinfo{year}{2020}).

\bibitem{ye2019improvement}
\bibinfo{author}{Ye, W.} \emph{et~al.}
\newblock \bibinfo{title}{Improvement of self-referenced continuous-variable quantum key distribution with quantum photon catalysis}.
\newblock \emph{\bibinfo{journal}{Opt. Express}} \textbf{\bibinfo{volume}{27}}, \bibinfo{pages}{17186--17198} (\bibinfo{year}{2019}).

\bibitem{hu2021performance}
\bibinfo{author}{Hu, J.}, \bibinfo{author}{Liao, Q.}, \bibinfo{author}{Mao, Y.} \& \bibinfo{author}{Guo, Y.}
\newblock \bibinfo{title}{Performance improvement of unidimensional continuous-variable quantum key distribution using zero-photon quantum catalysis}.
\newblock \emph{\bibinfo{journal}{Quantum Inf. Process.}} \textbf{\bibinfo{volume}{20}}, \bibinfo{pages}{1--20} (\bibinfo{year}{2021}).

\bibitem{walk2013security}
\bibinfo{author}{Walk, N.}, \bibinfo{author}{Ralph, T.~C.}, \bibinfo{author}{Symul, T.} \& \bibinfo{author}{Lam, P.~K.}
\newblock \bibinfo{title}{Security of continuous-variable quantum cryptography with gaussian postselection}.
\newblock \emph{\bibinfo{journal}{Phys. Rev. A}} \textbf{\bibinfo{volume}{87}}, \bibinfo{pages}{020303} (\bibinfo{year}{2013}).

\bibitem{fiuravsek2012gaussian}
\bibinfo{author}{Fiur{\'a}{\v{s}}ek, J.} \& \bibinfo{author}{Cerf, N.~J.}
\newblock \bibinfo{title}{Gaussian postselection and virtual noiseless amplification in continuous-variable quantum key distribution}.
\newblock \emph{\bibinfo{journal}{Phys. Rev. A}} \textbf{\bibinfo{volume}{86}}, \bibinfo{pages}{060302} (\bibinfo{year}{2012}).

\bibitem{hosseinidehaj2020finite}
\bibinfo{author}{Hosseinidehaj, N.}, \bibinfo{author}{Lance, A.~M.}, \bibinfo{author}{Symul, T.}, \bibinfo{author}{Walk, N.} \& \bibinfo{author}{Ralph, T.~C.}
\newblock \bibinfo{title}{Finite-size effects in continuous-variable quantum key distribution with gaussian postselection}.
\newblock \emph{\bibinfo{journal}{Phys. Rev. A}} \textbf{\bibinfo{volume}{101}}, \bibinfo{pages}{052335} (\bibinfo{year}{2020}).

\bibitem{chrzanowski2014measurement}
\bibinfo{author}{Chrzanowski, H.~M.} \emph{et~al.}
\newblock \bibinfo{title}{Measurement-based noiseless linear amplification for quantum communication}.
\newblock \emph{\bibinfo{journal}{Nat. Photonics}} \textbf{\bibinfo{volume}{8}}, \bibinfo{pages}{333--338} (\bibinfo{year}{2014}).

\bibitem{zhao2017characterization}
\bibinfo{author}{Zhao, J.}, \bibinfo{author}{Haw, J.~Y.}, \bibinfo{author}{Symul, T.}, \bibinfo{author}{Lam, P.~K.} \& \bibinfo{author}{Assad, S.~M.}
\newblock \bibinfo{title}{Characterization of a measurement-based noiseless linear amplifier and its applications}.
\newblock \emph{\bibinfo{journal}{Phys. Rev. A}} \textbf{\bibinfo{volume}{96}}, \bibinfo{pages}{012319} (\bibinfo{year}{2017}).

\bibitem{zhao2023enhancing}
\bibinfo{author}{Zhao, J.} \emph{et~al.}
\newblock \bibinfo{title}{Enhancing quantum teleportation efficacy with noiseless linear amplification}.
\newblock \emph{\bibinfo{journal}{Nat. Commun.}} \textbf{\bibinfo{volume}{14}}, \bibinfo{pages}{4745} (\bibinfo{year}{2023}).

\bibitem{shajilal2024improving}
\bibinfo{author}{Shajilal, B.} \emph{et~al.}
\newblock \bibinfo{title}{Improving gaussian channel simulation using non-unity gain heralded quantum teleportation}.
\newblock \emph{\bibinfo{journal}{arXiv preprint arXiv:2408.08667}}  (\bibinfo{year}{2024}).

\bibitem{li2016non}
\bibinfo{author}{Li, Z.} \emph{et~al.}
\newblock \bibinfo{title}{Non-gaussian postselection and virtual photon subtraction in continuous-variable quantum key distribution}.
\newblock \emph{\bibinfo{journal}{Phys. Rev. A}} \textbf{\bibinfo{volume}{93}}, \bibinfo{pages}{012310} (\bibinfo{year}{2016}).

\bibitem{zhong2018self}
\bibinfo{author}{Zhong, H.} \emph{et~al.}
\newblock \bibinfo{title}{Enhancing of self-referenced continuous-variable quantum key distribution with virtual photon subtraction}.
\newblock \emph{\bibinfo{journal}{Entropy}} \textbf{\bibinfo{volume}{20}}, \bibinfo{pages}{578} (\bibinfo{year}{2018}).

\bibitem{zhong2020virtual}
\bibinfo{author}{Zhong, H.}, \bibinfo{author}{Guo, Y.}, \bibinfo{author}{Mao, Y.}, \bibinfo{author}{Ye, W.} \& \bibinfo{author}{Huang, D.}
\newblock \bibinfo{title}{Virtual zero-photon catalysis for improving continuous-variable quantum key distribution via gaussian post-selection}.
\newblock \emph{\bibinfo{journal}{Sci. Rep.}} \textbf{\bibinfo{volume}{10}}, \bibinfo{pages}{17526} (\bibinfo{year}{2020}).

\bibitem{garcia2006unconditional}
\bibinfo{author}{Garc{\'\i}a-Patr{\'o}n, R.} \& \bibinfo{author}{Cerf, N.~J.}
\newblock \bibinfo{title}{Unconditional optimality of gaussian attacks against continuous-variable quantum key distribution}.
\newblock \emph{\bibinfo{journal}{Phys. Rev. Lett.}} \textbf{\bibinfo{volume}{97}}, \bibinfo{pages}{190503} (\bibinfo{year}{2006}).

\bibitem{wolf2006extremality}
\bibinfo{author}{Wolf, M.~M.}, \bibinfo{author}{Giedke, G.} \& \bibinfo{author}{Cirac, J.~I.}
\newblock \bibinfo{title}{Extremality of gaussian quantum states}.
\newblock \emph{\bibinfo{journal}{Phys. Rev. Lett.}} \textbf{\bibinfo{volume}{96}}, \bibinfo{pages}{080502} (\bibinfo{year}{2006}).

\bibitem{holevo1998capacity}
\bibinfo{author}{Holevo, A.~S.}
\newblock \bibinfo{title}{The capacity of the quantum channel with general signal states}.
\newblock \emph{\bibinfo{journal}{IEEE Trans. Inf. Theory}} \textbf{\bibinfo{volume}{44}}, \bibinfo{pages}{269--273} (\bibinfo{year}{1998}).

\bibitem{weedbrook2012gaussian}
\bibinfo{author}{Weedbrook, C.} \emph{et~al.}
\newblock \bibinfo{title}{Gaussian quantum information}.
\newblock \emph{\bibinfo{journal}{Rev. Mod. Phys.}} \textbf{\bibinfo{volume}{84}}, \bibinfo{pages}{621--669} (\bibinfo{year}{2012}).

\bibitem{weedbrook2012continuous}
\bibinfo{author}{Weedbrook, C.}, \bibinfo{author}{Pirandola, S.} \& \bibinfo{author}{Ralph, T.~C.}
\newblock \bibinfo{title}{Continuous-variable quantum key distribution using thermal states}.
\newblock \emph{\bibinfo{journal}{Phys. Rev. A}} \textbf{\bibinfo{volume}{86}}, \bibinfo{pages}{022318} (\bibinfo{year}{2012}).

\bibitem{li2023continuous}
\bibinfo{author}{Li, L.} \emph{et~al.}
\newblock \bibinfo{title}{Continuous-variable quantum key distribution with on-chip light sources}.
\newblock \emph{\bibinfo{journal}{Photonics Res.}} \textbf{\bibinfo{volume}{11}}, \bibinfo{pages}{504--516} (\bibinfo{year}{2023}).

\bibitem{sayat2024satellite}
\bibinfo{author}{Sayat, M.} \emph{et~al.}
\newblock \bibinfo{title}{Satellite-to-ground continuous variable quantum key distribution: The gaussian and discrete modulated protocols in low earth orbit}.
\newblock \emph{\bibinfo{journal}{IEEE Trans. Commun.}}  (\bibinfo{year}{2024}).

\bibitem{eisert2002distilling}
\bibinfo{author}{Eisert, J.}, \bibinfo{author}{Scheel, S.} \& \bibinfo{author}{Plenio, M.~B.}
\newblock \bibinfo{title}{Distilling gaussian states with gaussian operations is impossible}.
\newblock \emph{\bibinfo{journal}{Phys. Rev. Lett.}} \textbf{\bibinfo{volume}{89}}, \bibinfo{pages}{137903} (\bibinfo{year}{2002}).

\bibitem{denys2021explicit}
\bibinfo{author}{Denys, A.}, \bibinfo{author}{Brown, P.} \& \bibinfo{author}{Leverrier, A.}
\newblock \bibinfo{title}{Explicit asymptotic secret key rate of continuous-variable quantum key distribution with an arbitrary modulation}.
\newblock \emph{\bibinfo{journal}{Quantum}} \textbf{\bibinfo{volume}{5}}, \bibinfo{pages}{540} (\bibinfo{year}{2021}).

\bibitem{leverrier2009unconditional}
\bibinfo{author}{Leverrier, A.} \& \bibinfo{author}{Grangier, P.}
\newblock \bibinfo{title}{Unconditional security proof of long-distance continuous-variable quantum key distribution with discrete modulation}.
\newblock \emph{\bibinfo{journal}{Phys. Rev. Lett.}} \textbf{\bibinfo{volume}{102}}, \bibinfo{pages}{180504} (\bibinfo{year}{2009}).

\bibitem{ghorai2019asymptotic}
\bibinfo{author}{Ghorai, S.}, \bibinfo{author}{Grangier, P.}, \bibinfo{author}{Diamanti, E.} \& \bibinfo{author}{Leverrier, A.}
\newblock \bibinfo{title}{Asymptotic security of continuous-variable quantum key distribution with a discrete modulation}.
\newblock \emph{\bibinfo{journal}{Phys. Rev. X}} \textbf{\bibinfo{volume}{9}}, \bibinfo{pages}{021059} (\bibinfo{year}{2019}).

\bibitem{tian2023high}
\bibinfo{author}{Tian, Y.} \emph{et~al.}
\newblock \bibinfo{title}{High-performance long-distance discrete-modulation continuous-variable quantum key distribution}.
\newblock \emph{\bibinfo{journal}{Opt. Lett.}} \textbf{\bibinfo{volume}{48}}, \bibinfo{pages}{2953--2956} (\bibinfo{year}{2023}).

\bibitem{kanitschar2022optimizing}
\bibinfo{author}{Kanitschar, F.} \& \bibinfo{author}{Pacher, C.}
\newblock \bibinfo{title}{Optimizing continuous-variable quantum key distribution with phase-shift keying modulation and postselection}.
\newblock \emph{\bibinfo{journal}{Physical Review Applied}} \textbf{\bibinfo{volume}{18}}, \bibinfo{pages}{034073} (\bibinfo{year}{2022}).

\bibitem{grosshans2003virtual}
\bibinfo{author}{Grosshans, F.}, \bibinfo{author}{Cerf, N.~J.}, \bibinfo{author}{Wenger, J.}, \bibinfo{author}{Tualle-Brouri, R.} \& \bibinfo{author}{Grangier, P.}
\newblock \bibinfo{title}{Virtual entanglement and reconciliation protocols for quantum cryptography with continuous variables}.
\newblock \emph{\bibinfo{journal}{arXiv preprint quant-ph/0306141}}  (\bibinfo{year}{2003}).

\bibitem{laudenbach2018continuous}
\bibinfo{author}{Laudenbach, F.} \emph{et~al.}
\newblock \bibinfo{title}{Continuous-variable quantum key distribution with gaussian modulation—the theory of practical implementations}.
\newblock \emph{\bibinfo{journal}{Adv. Quantum Technol.}} \textbf{\bibinfo{volume}{1}}, \bibinfo{pages}{1800011} (\bibinfo{year}{2018}).

\bibitem{grosshans2002reverse}
\bibinfo{author}{Grosshans, F.} \& \bibinfo{author}{Grangier, P.}
\newblock \bibinfo{title}{Reverse reconciliation protocols for quantum cryptography with continuous variables}.
\newblock \emph{\bibinfo{journal}{arXiv preprint quant-ph/0204127}}  (\bibinfo{year}{2002}).

\bibitem{williamson1936algebraic}
\bibinfo{author}{Williamson, J.}
\newblock \bibinfo{title}{On the algebraic problem concerning the normal forms of linear dynamical systems}.
\newblock \emph{\bibinfo{journal}{Am. J. Math.}} \textbf{\bibinfo{volume}{58}}, \bibinfo{pages}{141--163} (\bibinfo{year}{1936}).

\bibitem{cariolaro2016reexamination}
\bibinfo{author}{Cariolaro, G.} \& \bibinfo{author}{Pierobon, G.}
\newblock \bibinfo{title}{Reexamination of bloch-messiah reduction}.
\newblock \emph{\bibinfo{journal}{Phys. Rev. A}} \textbf{\bibinfo{volume}{93}}, \bibinfo{pages}{062115} (\bibinfo{year}{2016}).

\bibitem{lu2022micius}
\bibinfo{author}{Lu, C.-Y.}, \bibinfo{author}{Cao, Y.}, \bibinfo{author}{Peng, C.-Z.} \& \bibinfo{author}{Pan, J.-W.}
\newblock \bibinfo{title}{Micius quantum experiments in space}.
\newblock \emph{\bibinfo{journal}{Reviews of Modern Physics}} \textbf{\bibinfo{volume}{94}}, \bibinfo{pages}{035001} (\bibinfo{year}{2022}).

\end{thebibliography}

\bibliographystyle{naturemag}

\clearpage
\twocolumngrid 
\onecolumngrid 
\appendix
\section*{Supplementary Material}

\section{\label{sec:experimental_parameters}Detailed Description of the Experimental Parameters}
In this section, \ZJ{we outline the procedure to implement data analysis in our experiments. Table~\ref{tab:experimental_parameters} summarises the experimental parameters for the operational regimes shown in Fig.(3) of the main text}. For each regime, the reconciliation efficiency is set to $\beta = 92\%$.
\begin{table*}[hbt!]
\caption{\label{tab:experimental_parameters}
Experimental parameters for each regime.%
\footnote{Experimental imperfections cause asymmetry in the $x$ and $p$ quadratures. The first row shows values for $x$, the second for $p$. Empty cells indicate that post-selection was not required. Distance assumes a fibre loss of 0.2 dB/km. $V_{\text{mod}}$: modulation variance in shot noise units. $W$: Eve's variance as $W_x$ and $W_p$ for $x$ and $p$ quadratures.}
}
\centering
\begin{tabular}{|c|c|c|c|c|c|c|}
\hhline{|=|=|=|=|=|=|=|}
\makecell{Distance\\(km)} &
\makecell{Transmission} &
\makecell{$V_{\text{mod}}$\\(SNU)} &
\makecell{$W_{x/p}$\\(SNU)} &
\makecell{Key rate\\(bits/use)} &
\makecell{After Alice\\(bits/use)} &
\makecell{After Bob\\(bits/use)} \\
\hline
\makecell{6.03$\pm$0.006} &
\makecell{0.757$\pm$0.0002} &
\makecell{12.62$\pm$0.009\\13.52$\pm$0.009} &
\makecell{1.02$\pm$0.004\\1.04$\pm$0.004} &
\makecell{0.4839$\pm$0.005} &
-- & -- \\
\hline
\makecell{15.11$\pm$0.007} &
\makecell{0.499$\pm$0.0002} &
\makecell{12.65$\pm$0.009\\13.63$\pm$0.010} &
\makecell{1.02$\pm$0.002\\1.03$\pm$0.002} &
\makecell{0.1433$\pm$0.0028} &
-- & -- \\
\hline
\makecell{29.14$\pm$0.008} &
\makecell{0.261$\pm$0.0001} &
\makecell{12.74$\pm$0.009\\13.52$\pm$0.009} &
\makecell{1.02$\pm$0.001\\1.01$\pm$0.001} &
\makecell{$-0.0093\pm$0.0029} &
\makecell{0.0025$\pm$0.0012} &
\makecell{0.0038$\pm$0.0011} \\
\hline
\makecell{39.06$\pm$0.010} &
\makecell{0.166$\pm$0.0001} &
\makecell{12.58$\pm$0.008\\13.40$\pm$0.009} &
\makecell{1.00$\pm$0.001\\1.00$\pm$0.001} &
\makecell{0.0099$\pm$0.0029} &
\makecell{0.0169$\pm$0.0020} &
-- \\
\hhline{|=|=|=|=|=|=|=|}
\end{tabular}
\end{table*}

\ZJ{Key to the data analysis is converting from the prepare-and-measure (PM) scheme to the entanglement-based (EB) scheme. This conversion enables the estimation of Eve's information, which in turn allows for calculating the final key rates.} \ozlem{To this end, we begin by expressing Bob's raw data after detection, following the experimental setup illustrated in Fig. 1(a) of the main text
\begin{equation}
    \label{eq:bobs_raw_data}
    x_b^r=\sqrt{T}(g_{e_x} x_a^r+x_{t}^r+x_{SN}^r)+\sqrt{1-T}x_{E_1}^r,
\end{equation}
where $g_{e_x}$ is the electronic gain of the system and $x_a^r$ represents Alice's raw data, generated by sampling the output of function generators, while $x_{SN}^r$, $x_{E_1}^r$, and $x_t^r$ correspond to shot noise, channel noise (introduced by Eve), and trusted excess noise, respectively. Alice and Bob's raw data are then calibrated to shot noise units as follows}
\begin{subequations}
    \begin{equation}
        x_a = \frac{x_a^r}{\sqrt{V_{SN} - V_{DN}}},
        \label{eq:xa_rescaled}
    \end{equation}
    \begin{equation}
        x_b = \frac{x_b^r}{\sqrt{V_{SN} - V_{DN}}},
        \label{eq:xb_rescaled}
    \end{equation}
\end{subequations}
where $V_{SN}$ and $V_{DN}$ represent the variances of the shot noise and dark noise respectively. \ozlem{After the calibration, Bob's data becomes
\begin{equation}
    \label{eq:bobs_raw_data_calibrated}
    x_b=\sqrt{T}(g_{e_x} x_a+x_{t}+x_{SN})+\sqrt{1-T}x_{E_1}.
\end{equation}
Typically, dark noise is subtracted from the variance of the raw samples. However, given our high dark noise clearance of 20 dB, this correction was not applied. After calibrating the shot noise, the initial PM covariance matrix between Alice and Bob for the $x$ quadrature can be calculated as follows}
\begin{equation}
    \sigma_{AB_x}^{PM}=\begin{pmatrix}
        V_{a_x} & C_{ab_x} \\
        C_{ab_x} & V_{b_x}
    \end{pmatrix},
    \label{eq:initial_pm_cm}
\end{equation}
where $V_{a_x}$ represents Alice's variance \ZJ{$\langle \Delta x_{a} \Delta x_a \rangle$}, $C_{ab_x}$ denotes the covariance between Alice and Bob \ozlem{$\langle \Delta x_{a} \Delta x_{b} \rangle=g_{e_x}\sqrt{T}V_{a_x}$}, and $V_{b_x}$ is Bob's variance of the homodyne measurement for the $x$ quadrature \ZJ{$\langle \Delta x_{b} \Delta x_{b} \rangle$}. 

\ZJ{After the shot noise calibration, Alice renormalises her data by the electronic gain of the system, $g_{e_x}$, to reveal the physical states she prepared. The gain is determined by setting} the channel transmission to $1$, corresponding to no loss, as
\begin{equation}
    g_{e_x}=\frac{C_{ab_x}}{V_{a_x}}.
    \label{eq:gain}
\end{equation}
\ozlem{Notably, this gain also accounts for Bob's homodyne detection efficiencies ($\sim$ 97.6\%), where we have taken into account the quantum efficiency of the photodiodes ($\sim$ 99\%), mode-matching visibility ($\sim$ 99.3\%).} The calculated gains for our experiment \ZJ{are} $g_{e_x}=8.33\pm0.01$ and $g_{e_p}=19.59\pm0.02$ for \ZJ{the amplitude and phase quadratures, respectively}. Using these gains, rescaled covariance matrix \ZJ{for the $x$ quadrature} in the PM scheme becomes
\begin{equation}
    \sigma_{AB_x}^{PM'}=\begin{pmatrix}
        g_{e_x}^2V_{a_x} & g_{e_x}C_{ab_x} \\
        g_{e_x}C_{ab_x} & V_{b_x}
    \end{pmatrix}=\begin{pmatrix}
        V_{mod_x} & C_{ab_x}' \\
        C_{ab_x}' & V_{b_x}
    \end{pmatrix},
    \label{eq:rescaled_pm_cm}
\end{equation}
where \ozlem{$C_{ab_x}'=g_{e_x}^2\sqrt{T}V_{a_x}=\sqrt{T}V_{mod_x}$ and $V_{mod_x}$ represents the rescaled modulation variance of the $x$ quadrature}. \ZJ{Similarly, $\sigma_{AB_p}^{PM'}$ is also obtained from the expression above for the phase quadrature.} Note that Bob's homodyne variance, $V_{b_x}$, does not need to be scaled as it already includes the system gain when the measurement is performed. \ozlem{Therefore, Bob's variance becomes
\begin{equation}
    \label{eq:bobs_variance_in_shot_noise}
    \langle \Delta x_b \Delta x_b \rangle=T(g_{e_x}^2 V_{a_x}+\xi_x+1)+(1-T)W_x=T(V_{mod_x}+\xi_x+1)+(1-T)W_x.
\end{equation}
Here, $W_x$ denotes Eve's variance while $\xi_x$ represents the variance of the trusted excess noise.} In an ideal scenario with no system loss, Bob's variance should be $V_{b_x}=V_{mod_x}+1$. However, due to noise in the state preparation during our experiment, the variance after the modulators becomes $V_{mod_x}+\xi_{x}$. Therefore, we model Bob's variance as $V_{b_x}=T(V_{mod_x}+1+\xi_x)+(1-T)W_x$ and when $T=1$, the trusted excess noise, $\xi_x$ and $\xi_p$ for the $x$ and $p$ quadratures \ZJ{can be estimated from} 
\begin{subequations}
    \begin{equation}
        \xi_x=V_{b_x}-V_{mod_x}-1,
    \end{equation}
    \begin{equation}
        \xi_p=V_{b_p}-V_{mod_p}-1.
    \end{equation}
\end{subequations}
The experimental values obtained for these trusted noises are $\xi_x = 0.28 \pm 0.003$~SNU and $\xi_p = 0.40 \pm 0.002$~SNU. 

For the security analysis, we use the EB scheme~\cite{weedbrook2012gaussian,grosshans2003virtual}. For each regime, the EB covariance matrix is obtained by rescaling the PM matrix given in Eq.~\eqref{eq:rescaled_pm_cm} following the method described in Laudenbach \textit{et al.}~\cite{laudenbach2018continuous} as below
\begin{equation}
    \sigma_{AB}^{EB}=\begin{pmatrix}
        V_{mod_x}+1 & 0 & \sqrt{\frac{V_{mod_x}+2}{V_{mod_x}}}C_{ab_x}' & 0\\
        0 & V_{mod_p}+1 & 0 & -\sqrt{\frac{V_{mod_p}+2}{V_{mod_p}}}C_{ab_p}'\\
        \sqrt{\frac{V_{mod_x}+2}{V_{mod_x}}}C_{ab_x}' & 0 & V_{b_x} & 0 \\
        0 & -\sqrt{\frac{V_{mod_p}+2}{V_{mod_p}}}C_{ab_p}' & 0 & V_{b_p}
    \end{pmatrix}=\begin{pmatrix}
    V_x & 0 & C_x & 0 \\
    0 & V_p & 0 & -C_p \\
    C_x & 0 & V_{b_x} & 0 \\
    0 & -C_p & 0 & V_{b_p}
    \end{pmatrix},
    \label{eq:cm_eb_alice_and_bob}
\end{equation}
where $C_{ab_x}'$ and $C_{ab_p}'$ are equal to $\sqrt{T}V_{mod_x}$ and $\sqrt{T}V_{mod_p}$, respectively~\cite{laudenbach2018continuous}. \ozlem{As mentioned in the main text, $V_x=V_{mod_x}+1$ and $C_x =\sqrt{T(V_x^2-1)}$. For the phase quadrature, the same equations apply with $V_{mod_x}$, $W_x$ and $\xi_x$ replaced by $V_{mod_p}$, $W_p$ and $\xi_p$, respectively. Additionally, when post-selection is applied, $V_{mod_x}$ and $V_{mod_p}$ must be replaced with $\Tilde{V}_{mod_x}$ and $\Tilde{V}_{mod_p}$, respectively.}

\section{\label{sec:theory_heterodyne}Theoretical Model of Post-selection with Heterodyne Detection}
\ZJ{In this section, we present the theoretical model of our protocol when heterodyne measurement is implemented at Bob's station, complementing the Methods section in the main text, which outlines the theory for homodyne measurement by Bob. }
\subsection{\label{sec:alices_postselection}Alice's Post-selection}
\ZJ{In this case, Alice's data and her post-selection remains unchanged. Therefore,} Eqs. (6) to (8) in the main text still hold. However, \ozlem{Bob's data, and therefore the calculation of the secret key, are different.} Because Bob performs heterodyne detection, there is \ZJ{an} added vacuum noise. Additionally, \ZJ{our theory model also} accounts for detector noise and efficiency, \ZJ{necessary to analyse the data collected from Data61 that is presented in Fig.~(4) of the main text. Bob's measured variance can be expressed as} 
\begin{equation}
    V_{b_x}^m=\frac{\eta T (V_{mod_x}+1+\xi_x)+\eta (1-T) W_x + (1-\eta)+1}{2} + \xi_d.
    \label{eq:bobs_heterodyne_variance}
\end{equation}
Here, $\xi_x$ is the trusted excess noise attributed to the state preparation, while $W_x$ is Eve's thermal variance in the $x$ quadrature. $\eta$ represents the detector efficiency and $\xi_d$ is the electronic noise of the detector. Bob's variance in the $p$ quadrature is the same except variables $V_{mod_x}$, $\xi_x$ and $W_x$ are now replaced with $V_{mod_p}$, $\xi_p$ and $W_p$. Note that variance of \ZJ{the quantum state received by Bob} is $V_{b_x}=2V_{b_x}^m-1$ \ZJ{given by}
\begin{equation}
    V_{b_x}=\eta T (V_{mod_x}+1+\xi_x)+\eta (1-T) W_x + (1-\eta) + 2\xi_d.
    \label{eq:bobs_heterodyne_variance_new}
\end{equation}
Similarly, the covariance between Alice and Bob going from PM scheme to EB scheme needs to be scaled, \ozlem{where the initial PM covariance matrix (already normalised to the shot noise units) for the heterodyne detection is expressed as
\begin{equation}
    \sigma_{AB_x}^{PM}=\begin{pmatrix}
        V_{mod_x} & C_{ab_x}' \\
        C_{ab_x}' & V_{b_x}^{m}
    \end{pmatrix},
    \label{eq:initial_pm_cm_ch}
\end{equation}
where the covariance term $C_{ab_x}'$ is given by  $C_{ab_x}' = \sqrt{\eta T/2} V_{mod_x}$. Then the EB representation of the covariance matrix becomes}
\begin{equation}
    \sigma_{AB}^{EB}=\begin{pmatrix}
        V_{mod_x}+1 & 0 & \sqrt{\frac{2(V_{mod_x}+2)}{V_{mod_x}}}C_{ab_x}' & 0\\
        0 & V_{mod_p}+1 & 0 & -\sqrt{\frac{2(V_{mod_p}+2)}{V_{mod_p}}}C_{ab_p}'\\
        \sqrt{\frac{2(V_{mod_x}+2)}{V_{mod_x}}}C_{ab_x}' & 0 & 2V_{b_x}^m-1 & 0 \\
        0 & -\sqrt{\frac{2(V_{mod_p}+2)}{V_{mod_p}}}C_{ab_p}' & 0 & 2V_{b_p}^m-1
    \end{pmatrix}=\begin{pmatrix}
    V_x & 0 & C_x & 0 \\
    0 & V_p & 0 & -C_p \\
    C_x & 0 & V_{b_x} & 0 \\
    0 & -C_p & 0 & V_{b_p}
    \end{pmatrix}.
    \label{eq:cm_eb_alice_and_bob_heterodyne}
\end{equation}
When Alice performs post-selection, all instances of $V_{mod_x}$ are replaced by the effective modulation variance $\Tilde{V}_{mod_x}$, as calculated using Eq.~(8) of the main text. Additionally, $C_x = \sqrt{\eta T(V_x^2 - 1)}$, where $V_x = V_{mod_x} + 1$ without post-selection, and $V_x = \Tilde{V}_{mod_x} + 1$ when Alice performs Gaussian post-selection.

As Bob performs heterodyne detection, the mutual information is different from the homodyne case. Given that Alice prepares a two-mode squeezed vacuum state in the equivalent EB scenario and sends one arm to Bob~\cite{grosshans2003virtual,garcia2007quantum} while measuring the other with heterodyne detection, her variance of the classical data is $V_{A_x}=(V_x+1)/2$. Bob's heterodyne variance is also $V_{b_x}^m=(V_{b_x}+1)/2$. Since both Bob and Alice perform heterodyne detection in the equivalent EB representation, the covariance terms scales by $1/2$ and becomes $C_x/2$. Therefore, $V_{A_x|B_x}=V_{A_x}-C_x^2/(4V_{b_x}^m)$. The mutual information is then calculated using Eq. (12) of the main text for $x$ and $p$ quadrature separately and averaged over both quadratures with $I_{AB}=0.5I_{AB_x}+0.5I_{AB_p}$. 

Similar to the homodyne case, when Alice performs a post-selection, Eve's average and conditional \ozlem{covariance} matrices need to scale as the effective modulation variance changes. Eve's average covariance matrix follow Eqs. (13) and (14) of the main text. However, her conditional covariance matrix now depends on measurement outcomes of both quadratures instead of one. Therefore, the conditional covariance matrix can be calculated from~\cite{laudenbach2018continuous}
\begin{equation}
    \label{eq:eves_conditional_state_heterodyne}
    \sigma_{E_1E_2|B}=\sigma_{E_1E_2}-\frac{1}{V_{b_x}+1}\sigma_c\sigma_c^T,
\end{equation}
where $\sigma_{E_1E_2}$ is Eve's average covariance matrix from Eq.~(14) of the main text and $\sigma_c$ is the matrix describing the covariance terms between Eve and Bob given as
\begin{equation}
    \sigma_c=\begin{pmatrix}
        \sqrt{\eta T(1-T)}\big(W_x-(V_x+\xi_x)\big) & 0 \\
        0 & \sqrt{\eta T(1-T)}\big(W_p-(V_p+\xi_p)\big) \\
        \sqrt{\eta (1-T)}C_{E_x} & 0 \\
        0 & \sqrt{\eta (1-T)}C_{E_p}
    \end{pmatrix},
    \label{eq:sigmac_heterodyne}
\end{equation}
where $C_{E_x}$ and $C_{E_p}$ are expressed in the Methods section of the main text. For full derivation of the matrix $\sigma_c$, refer to Laudenbach \textit{et al.}~\cite{laudenbach2018continuous}. 

Eve's information is bounded by the Holevo bound~\cite{holevo1998capacity} using Eq.~(15) of the main text. The secret key rate is then calculated using 
\begin{equation}
    K=P_{A_x}P_{A_p}(\beta I_{AB}-I_{E}),
    \label{eq:keyrate_heterodyne_alice}
\end{equation}
where $P_{A_x}$ where  $P_{A_p}$ are the probability of success of Alice's post-selection calculated using Eq.~(7) of the main text. Note that the total probability of success in the heterodyne case is $P_{A_t}=P_{A_x}\times P_{A_p}$.

\subsection{\label{sec:bobs_postselection}Bob's Post-selection}
The filter function that Bob uses for heterodyne measurements is given as
\begin{equation}
    F_B(x_b,p_b) =
    \begin{cases}
	0 & \text{$-c^2<x_{b}^2+p_b^2<c^2$} \\ 
	1 & \text{$\mathrm{elsewhere}$},
    \end{cases}
    \label{eq:bobs_filter_heterodyne}
\end{equation}
where $c$ represents the cut-off parameter of the filter which can take the values from $0$ to the largest measurement outcome of Bob. Bob's distribution before the post-selection can be expressed as
\begin{equation}
    \label{eq:bobs_dist_heterodyne}
    p_b(x_b,p_b)=\frac{1}{\pi\sqrt{(V_{b_x}+1)(V_{b_p}+1)}}\mathrm{exp}\bigg[-\frac{x_b^2}{(V_{b_x}+1)}-\frac{p_b^2}{(V_{b_p}+1)}\bigg].
\end{equation}
After Bob's post-selection process, the probability of success for a given cut-off can be found from 
\begin{equation}
    P_B(c)=\int_{-\infty}^\infty \int_{-\infty}^\infty F_B(x_b,p_b)p_b(x_b,p_b) \, dx_b dp_b.
\end{equation}
Notably, if $V_{b_x}=V_{b_p}$, the probability of success simplifies to $P_{B}(c)=\exp\big[-\frac{c^2}{V_{b_x}+1}\big]$. Bob's output distribution after the post-selection process becomes
\begin{equation}
    \Tilde{p}_b(x_b,p_b)=\begin{cases}
        0 & \text{$-c^2<x_b^2+p_b^2<c^2$} \\
        \frac{p_b(x_b,p_b)}{P_{B}(c)} & \text{$\mathrm{elsewhere}$}.
    \end{cases}
    \label{eq:bobs_output_distribuion_heterodyne}
\end{equation}

Similar to the homodyne case, we avoid using the Gaussian extremality and instead bound Eve's information directly from her density matrices. As described in the Methods section of the main text, we use Williamson's decomposition~\cite{williamson1936algebraic} to obtain a symplectic matrix, $s$, and a diagonal matrix, $\omega$. We then further decompose the symplectic matrix, $s$ using Bloch-Messiah decomposition~\cite{cariolaro2016reexamination} to express $s$ as a series of beamsplitters, squeezers and phase rotations: $s=s_us_is_v$ in the covariance matrix notation and $S=S_US_IS_V$ in the density matrix notation. In our case for Data61's results, both $s_u$ and $s_v$ are both expressed through phase rotations and beamsplitters: $S_U=R(\theta_1, \theta_2) B(\tau_U) R(\theta_3, \theta_4)$ and $S_V=R(\theta_5, \theta_6) B(\tau_V) R(\theta_7, \theta_8)$, where each parameter needs to be solved. The $s_i$ component is expressed through \ozlem{two-mode} squeezing operations, $S_I=S(r_1, r_2)$. Therefore, Eve's overall conditional density matrix is written in terms of the symplectic density matrix $S$ and a mixed thermal state $\Lambda$
\begin{equation}
    \label{eq:eves_density_matrix_hetedrodyne}
    \rho_{E_{1}E_{2}|B}=S \Lambda S^\dagger.
\end{equation}
Note that $\rho_{E_{1}E_{2}|B}$ in Eq.~\eqref{eq:eves_density_matrix_hetedrodyne} does not provide any information about Bob's measurement outcomes, therefore, we assume that the conditional matrix is centred at zero. We then apply a displacement operator on the conditional matrix centred at zero to determine Eve's conditional density matrix for each measurement outcome Bob gets.
\begin{equation}
    \rho_{E_{1}E_{2}|B}(x_b,p_b)=D(\alpha_1(x_{b},p_b),\alpha_2(x_{b},p_b))\rho_{E_{1}E_{2}|B}(x_b=0,p_b=0)D^\dagger(\alpha_1(x_{b},p_b),\alpha_2(x_{b},p_b)), 
\end{equation}
where $D(\alpha_1(x_b,p_b),\alpha_2(x_b,p_b))$ represents the displacement operator for Eve's modes 1 and 2 in the form of 
\begin{equation}
    \label{eq:displacement_operator_ch}
    D(\alpha_1(x_{b},p_b),\alpha_2(x_{b},p_b))=\exp(\alpha_1(x_b,p_b)a^\dagger-\alpha_1(x_b,p_b)a)\otimes\exp(\alpha_2(x_b,p_b)a^\dagger-\alpha_2(x_b,p_b)a),
\end{equation}
where $\alpha_1$ and $\alpha_2$ are functions of Eve's conditional means based on Bob's measurement outcomes, $x_b$ and $p_b$. These conditional means for Eve are calculated from
\begin{equation}
\label{eq:conditional_means_homodyne2}
    \mu_{E_1E_2}(x_b,p_b)=\frac{\sigma_c}{\sqrt{2}}\begin{pmatrix}
        V_{b_x}^m & 0\\
        0 & V_{b_p}^m
    \end{pmatrix}^{-1}
    \begin{pmatrix}
       x_b \\ p_b
    \end{pmatrix}.
\end{equation}
The values of $\alpha_1$ and $\alpha_2$ are calculated from $\alpha_1(x_b,p_b)=(\mu_{E_1}(x_b)+i\mu_{E_1}(p_b))/2$ and $\alpha_2(x_b,p_b)=(\mu_{E_2}(x_b)+i\mu_{E_2}(p_b))/2$ where $\mu_{E_1}(x_b)$ , $\mu_{E_1}(p_b)$, $\mu_{E_2}(x_b)$ and $\mu_{E_2}(p_b)$ correspond to rows 1 to 4 of $\mu_{E_1E_2}(x_b,p_b)$, respectively.

Even though each conditional state has the same entropy, they contribute differently to Eve's average state due to the varying probabilities of obtaining each conditional state. To determine Eve's average density matrix, we multiply each conditional state by its corresponding probability from Bob's post-selected probability distribution and then sum them all. Bob's probability of obtaining a given outcome is calculated from 
\begin{equation}
    p_b'(x_i,p_j)=\int_{p_j-\Delta}^{p_j+\Delta}\int_{x_i-\Delta}^{x_i+\Delta} \Tilde{p}_b(x_i,p_j)\, dx_b dp_b,
    \label{eq:bobs_probability_pxi}
\end{equation}
where $\Delta$ is the width of the bins and in our results $\Delta\!=\!0.25$ for the heterodyne detection data. Summing over all the possible outcomes of Bob, Eve's average state becomes
\begin{equation}
    \rho_{E_1E_2}=\sum\limits_{p_{j}=-n_p}^{n_p} \sum\limits_{x_{i}=-n_x}^{n_x} p_b'(x_i,p_j) \rho_{E_{1}E_{2}|B}(x_i,p_j),
    \label{eq:eve_ave_denmat2}
\end{equation}
where $n_x$ and $n_p$ are the maximum number that Bob measures and $p_b'(x_i,p_j)$ denotes Bob’s probability of getting a given outcome, $x_i$ and $p_j$. Eve's information then can be calculated from
\begin{equation}
    I_{E}=S(\rho_{E_1E_2})-S(\rho_{E_{1}E_{2}|B}(0,0)),
    \label{eq:eves_information_density_matrix_heterodyne}
\end{equation}
where $S(\rho_{E_1E_2})$ represents the von Neumann entropy for Eve's average state while $S(\rho_{E_{1}E_{2}|B}(0,0))$ gives the von Neumann entropy of Eve's conditional state given Bob measures $x_b=0$ and $p_b=0$.

In order to calculate Alice and Bob's mutual information after Bob's post-selection, we fit Alice and Bob's data to the following multivariate Gaussian distribution
\begin{equation}
    f_{AB}(x_a,p_a,x_b,p_b)=\frac{1}{\sqrt{(2\pi)^k|\sigma_{AB}^{C}|}}\exp\bigg[-\frac{1}{2}v^T (\sigma_{AB}^{C})^{-1}v\bigg],
    \label{eq:multivariate_dist_Alice_Bob_heterodyne}
\end{equation}
where $|\sigma_{AB}^{C}|$ represents the determinant of the joint covariance matrix between Alice and Bob's classical variables. $v$ is a $k$-dimensional column vector given as $v=(x_a,p_a,x_b,p_b)$ and therefore, $k=4$. The classical covariance matrix between Alice and Bob is given as
\begin{equation}
    \sigma_{AB}^C=\begin{pmatrix}
        (V_x+1)/2 & 0 & C_x/2 & 0 \\
        0 & (V_x+1)/2 & 0 & -C_p/2 \\
        C_x/2 & 0 & (V_{b_x}+1)/2 & 0 \\
        0 & -C_p/2 & 0 & (V_{b_p}+1)/2
    \end{pmatrix}.
    \label{eq:alice_bob_cm_classical_heterodyne}
\end{equation}
Using Eq.~\eqref{eq:multivariate_dist_Alice_Bob_heterodyne}, we generate a probability table between Alice and Bob for each values of $x_a,x_b,p_a$ and $p_b$ where $x_a$ and $p_a$ range from $-14$ to $14$ with increments of $\Delta=0.25$, while $x_b$ and $p_b$ range from $-8$ to $8$ with increments of the same width. The mutual information is then calculated using
\begin{equation}
    I_{AB}=H_A+H_B-H_{AB},
\end{equation}
where the joint entropy between Alice and Bob, $H_{AB}$ is computed from
\begin{equation}
    H_{AB}=-\sum_{x_a=-n_a}^{n_a}\sum_{p_a=-n_a}^{n_a}\sum_{x_b=-c}^{c}\sum_{p_b=-\sqrt{c^2-x_b^2}}^{\sqrt{c^2-x_b^2}}\frac{f_{AB}(x_a,p_a,x_b,p_b)}{P_B(c)}\log_2\bigg[\frac{f_{AB}(x_a,p_a,x_b,p_b)}{P_B(c)}\bigg],
    \label{eq:mutual_information_hetedoyne_shannon}
\end{equation}
where $n_a$ represents the maximum number that Alice prepares. In order to compute $H_A$, we need to obtain Alice's probabilities after Bob's post-selection
\begin{equation}
    f_A(x_a,p_a)=\frac{1}{P_B(c)}\sum_{x_b=-c}^{c}\sum_{p_b=-\sqrt{c^2-x_b^2}}^{\sqrt{c^2-x_b^2}}f_{AB}(x_a,p_a,x_b,p_b).
    \label{eq:fa_heterodyne_bob}
\end{equation}
$H_A$ is then expressed as
\begin{equation}
    H_A=-\sum_{x_a=-n_a}^{n_a}\sum_{p_a=-n_a}^{n_a}f_A(x_a,p_a)\log_2[f_A(x_a,p_a)].
    \label{eq:ha_heterodyne_bob}
\end{equation}
Similarly, Bob's probability from the probability table after his post-selection becomes
\begin{equation}
    f_B(x_b,p_b)=\frac{1}{P_B(c)}\sum_{x_a=-n_a}^{n_a}\sum_{p_a=-n_a}^{n_a}f_{AB}(x_a,p_a,x_b,p_b).
    \label{eq:fb_heterodyne_bob}
\end{equation}
Using Eq.~\eqref{eq:fb_heterodyne_bob}, Bob's entropy is computed as
\begin{equation}
    H_B=-\sum_{x_b=-c}^{c}\sum_{p_b=-\sqrt{c^2-x_b^2}}^{\sqrt{c^2-x_b^2}}f_B(x_b,p_b)\log_2[f_B(x_b,p_b)].
    \label{eq:hb_heterodyne_bob}
\end{equation}

\subsection{Calculation of the Secret Key Rate}
When both Alice and Bob perform post-selection, the key rate is computed from
\begin{equation}
    K=P_BP_{A_x}P_{A_p}(\beta I_{AB} - I_E),
\end{equation}
where $P_B$ denotes probability of success of Bob's post-selection and, $P_{A_x}$ and $P_{A_p}$ are the probability of success of Alice's post-selection for the $x$ and $p$ quadratures, respectively.

\section{The Equivalence between the Entanglement-Based and Prepare-and-Measure Schemes for Alice's Post-selection}
\begin{figure*}[htp!]
\center{\includegraphics[width=\textwidth]
        {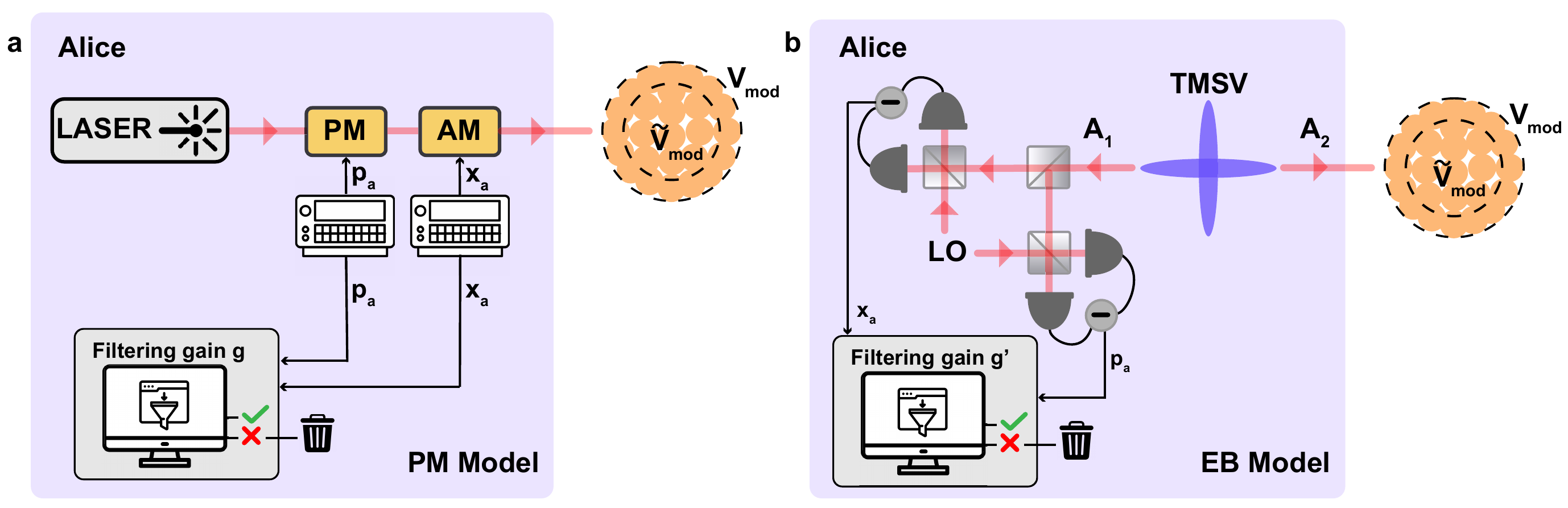}}
\caption{\label{fig:equivalence} The equivalence between the prepare-and-measure and entanglement-based models for Alice's post-selection. \textbf{(a)} Alice encodes coherent states $\ket{(x_a+ip_a)/2}$ from a Gaussian distribution with zero mean and variance, $V_{mod}$, using white noise generated by function generators. She samples $x_a$ and $p_a$ simultaneously. After data acquisition and once the states have been sent to Bob, Alice applies a Gaussian filter to her measurement outcomes with a filter gain, $g$. This shrinks the initial thermal state with a modulation variance of $V_{mod}$ to $\Tilde{V}_{mod}$. \textbf{(b)} Alice prepares a two-mode squeezed vacuum (TMSV) state with initial variance $V = V_{mod} + 1$ and zero mean. She performs heterodyne detection on mode $A_1$, followed by a Gaussian filter with gain $g'$. As a result, mode $A_2$ is projected onto a thermal state with zero mean and an effective modulation variance $\tilde{V}_{mod}$. 
}
\end{figure*}

\REVozlem{In this section, we derive the EB representation corresponding to Alice’s post-selection in the PM scheme as shown in Fig.~\ref{fig:equivalence}(a). Alternatively, one can begin with the EB model to construct an equivalent experimental description of Alice’s post-selection process. In Fig.~\ref{fig:equivalence}(b), Alice prepares a TMSV state and performs a heterodyne measurement on one mode, labeled $A_1$, while sending the other mode, $A_2$, to Bob. This measurement projects mode $A_2$ onto a thermal state with zero mean and variance $V$. The resulting covariance matrix after Alice’s measurement is given by
\begin{equation}
    \label{eq:alices_CM_A1A2}
    \sigma_{A_1A_2}^{EB}(V)=\begin{pmatrix}
        \frac{V+1}{2}\mathbb{I}_2 & \sqrt{\frac{V^2-1}{2}}\sigma_z\\
         \sqrt{\frac{V^2-1}{2}}\sigma_z & V\mathbb{I}_2.
    \end{pmatrix}
\end{equation}
For simplicity, we initially assume symmetric variance in both quadratures. The model is later extended to incorporate asymmetric quadrature variances. The joint Gaussian distribution corresponding to $\sigma_{A_1 A_2}^{\mathrm{EB}}$ is given by
\begin{equation}
\label{eq:alices_dist_A1A2}
    P_{A_1A_2}(V,x_a,p_a,x_b,p_b)=\frac{1}{\sqrt{4\pi^2|\sigma_{A_1A_2}^{EB}(V)|}}\exp\bigg[-\frac{1}{2}(x_a,p_a,x_b,p_b)\sigma_{A_1A_2}^{{EB}^{-1}}(V)(x_a,p_a,x_b,p_b)^T\bigg].
\end{equation}
When the quadratures are symmetric, Alice applies the following filter with an effective gain $g'$. This gain must be chosen to ensure consistency with the PM model derivation presented in the main manuscript. Therefore, Alice’s filter is given by
\begin{equation}
\label{eq:alices_filter_symmetric}
    F_{A}(x_a,p_a,g')=e^{-g'^2(x_a^2+p_a^2)}.
\end{equation}
Applying the filter on $A_1$, gives the following probability of success
\begin{equation}
    \label{eq:PS_symmetric}
    P_{A_t}(V,g')=\iiiint F_{A}(x_a,p_a,g')P_{A_1A_2}(V,x_a,p_a,x_b,p_b)\;d_{x_a}d_{p_a}d_{x_b}d_{p_b}=\frac{1}{g'^2(V+1)+1},
\end{equation}
where setting $g'=g\sqrt{\frac{2(V-1)}{(V+1)}}$ and $V-1=V_{mod}$ gives
\begin{equation}
\label{eq:ps_equal_A1A2}
    P_{A_t}(V_{mod},g)=\frac{1}{2g^2V_{mod}+1}.
\end{equation}
Recall that in the case of asymmetric quadratures, Alice's total success probability is given by $P_t = P_{A_x} \times P_{A_p}$, where $P_{A_x}$ and $P_{A_p}$ are defined in Eq.~(7) of the main manuscript. 
\begin{align}
\label{eq:joint_ps_prob}
    P_t(V_{mod_x},V_{mod_p},g_x,g_p)&=P_{A_x}(V_{mod_x},g_x)\times P_{A_p}(V_{mod_p},g_p)\nonumber \\
    &=\frac{1}{\sqrt{2g_x^2V_{mod_x}+1}}\times\frac{1}{\sqrt{2g_p^2V_{mod_p}+1}},
\end{align}
Setting $V_{mod_x}=V_{mod_p}=V_{mod}$ and $g_x=g_p=g$ gives the total probability of success as
\begin{equation}
    P_{t}(V_{mod},g)=\frac{1}{2g^2V_{mod}+1},
\end{equation}
which matches Eq.~\eqref{eq:ps_equal_A1A2}. After Alice's post-selection the output joint distribution is expressed as
\begin{equation}
\label{eq:output_dist_A1A2}
    \Tilde{P}_{A_1A_2}(V,x_a,p_a,x_b,p_b,g')=\frac{F_{A}(x_a,p_a,g')\times P_{A_1A_2}(V,x_a,p_a,x_b,p_b)}{P_{A_t}(V,g')}.
\end{equation}
Integrating over mode $A_1$ gives the distribution of $A_2$ as
\begin{equation}
    \label{eq:A2_dist}
    \Tilde{P}_{A_2}(V,x_b,p_b,g')=\iint \Tilde{P}_{A_1A_2}(V,x_a,p_a,x_b,p_b,g')\;d_{x_a}d_{p_a},
\end{equation}
which is used to determine the new thermal variance of $A_2$ after Alice's post-selection as
\begin{equation}
    \label{eq:new_thermal_variance_A2}
    \Tilde{V}(V,g')=\iint xb^2\Tilde{P}_{A_2}(V,x_b,p_b,g')d_{x_b}d_{p_b}=\frac{V+g'^2(V+1)}{g'^2(V+1)+1}.
\end{equation}
As $\Tilde{V}_{mod}=\Tilde{V}-1$ and $g'=g\sqrt{\frac{2(V-1)}{(V+1)}}$, the new modulation variance becomes
\begin{align}
    \label{eq:new_mod_A2}
    \Tilde{V}_{mod}(V,g')&=\Tilde{V}(V,g')-1 \nonumber \\
    &=\frac{V-1}{g'^2(V+1)+1}, \nonumber \\
    \Tilde{V}_{mod}(V_{mod},g)&=\frac{V_{mod}}{2g^2V_{mod}+1},
\end{align}
which is exactly the same as the modulation variance we derived in the main manuscript in Eq.(8).}

When the modulation variances are asymmetric due to experimental imperfections, this corresponds to Alice preparing a TMSV-like state in the EB scheme with unequal quadrature variances. Specifically, she generates two squeezed states with squeezing parameters $r_1$ and $r_2$ in the $x$ and $p$ quadratures, respectively. Upon performing a measurement on mode $A_1$, mode $A_2$ is projected into a thermal state with zero mean and variances $V_x$ and $V_p$ along the $x$ and $p$ quadratures. The effective covariance matrix then becomes
\begin{equation}
    \label{eq:alices_CM_A1A2_antisymmetric}
    \sigma_{A_1A_2}^{EB}(V_x,V_p)=\begin{pmatrix}
        \frac{V_x+1}{2} & 0 & \sqrt{\frac{V_x^2-1}{2}} & 0\\
        0 & \frac{V_p+1}{2} & 0 & -\sqrt{\frac{V_p^2-1}{2}} \\
        \sqrt{\frac{V_x^2-1}{2}} & 0 & V_x & 0 \\
        0 & -\sqrt{\frac{V_p^2-1}{2}} & 0 & V_p
    \end{pmatrix},
\end{equation}
where the joint distribution now depends on the variances $V_x$ and $V_p$, and retains the same functional form as Eq.~\eqref{eq:alices_dist_A1A2}. We denote the asymmetric distribution as $P_{A_1A_2}(V_x, V_p, x_a, p_a, x_b, p_b)$. Alice's filter, modified to accommodate this asymmetry, can be written as
\begin{equation}
\label{eq:alices_filter_asymmetric}
    F_{A}(x_a,p_a,g_x',g_p')=e^{-(g_x'^2x_a^2+g_p'^2p_a^2)},
\end{equation}
where $g_x'$ and $g_p'$ correspond to the effective gains applied to the $x$ and $p$ quadratures, respectively, in the asymmetric filtering scheme. The probability of success when Alice applies this filter to her measurement outcomes $x_a$ and $p_a$ becomes
\begin{align}
    \label{eq:ps_asymmetric_alice}
    P_{A_t}(V_x,V_p,g_x',g_p')&=\iiiint F_{A}(x_a,p_a,g_x',g_p')P_{A_1A_2}(V_x, V_p, x_a, p_a, x_b, p_b) \nonumber \\
    &=\frac{1}{\sqrt{(g_x'^2(V_x+1)+1)(g_p'^2(V_p+1)+1)}}.
\end{align}
By setting $g'_x = g_x \sqrt{\frac{2(V_x - 1)}{(V_x + 1)}}$, $g'_p = g_p \sqrt{\frac{2(V_p - 1)}{(V_p + 1)}}$, and defining the modulation variances as $V_{mod_x} = V_x - 1$ and $V_{mod_p} = V_p - 1$, the probability of success becomes
\begin{align}
\label{eq:final_ps_asymmetric}
    P_{A_t}(V_x,V_p,g_x,g_p)&=\frac{1}{\sqrt{(2g_x^2(V_x-1)+1)(2g_p^2(V_p-1)+1)}} \nonumber \\
     P_{A_t}(V_{mod_x},V_{mod_p},g_x,g_p)&=\frac{1}{\sqrt{(2g_x^2V_{mod_x}+1)(2g_p^2V_{mod_p}+1)}},
\end{align}
which is exactly the same as the PM model given in the main manuscript (see also Eq.~\eqref{eq:joint_ps_prob}). After the post-selection process, the distribution of $A_2$ becomes
\begin{equation}
\label{eq:anti_symmetric_A2}
    \Tilde{P}_{A_2}(V_x,V_p,x_b,p_b,g'_x,g'_p)=\iint \Tilde{P}_{A_1A_2}(V_x,V_p,x_a,p_a,x_b,p_b,g'_x,g'_p)\;d_{x_a}d_{p_a},
\end{equation}
then the thermal variance of $A_2$ is given by
\begin{equation}
    \label{eq:vnew_a2_asymmetric}
    \Tilde{V_x}(V_x,g_x')=\iint xb^2\Tilde{P}_{A_2}(V_x,V_p,x_b,p_b,g_x',g_p')d_{x_b}d_{p_b}=\frac{V_x+g_x'^2(V_x+1)}{g_x'^2(V_x+1)+1}.
\end{equation}
As $\Tilde{V}_{mod_x}=\Tilde{V_x}-1$ and $g_x' = g_x \sqrt{\frac{2(V_x - 1)}{(V_x + 1)}}$, the new modulation variance for the $x$ quadrature becomes
\begin{equation}
    \label{eq:vnew_a2_asymmetric2}
    \Tilde{V}_{mod_x}(V_x,g_x')=\frac{V_x-1}{2g_x^2(V_x-1)+1}=\frac{V_{mod_x}}{2g_x^2V_{mod_x}+1},
\end{equation}
which is equal to the modulation variance given in Eq.~(8) of the manuscript.

\section{\label{sec:free_space_parameters}Parameters of the Satellite-to-Ground Simulation and duty cycle estimation}
The results presented in Fig.~6 of the main text is calculated using the model described by Sayat \textit{et al.}~\cite{sayat2024satellite} where the parameters used in the model are described in Table~\ref{tab:free_space_parameters_table}.
\begin{figure*}[ht!]
\center{\includegraphics[width=\textwidth]
        {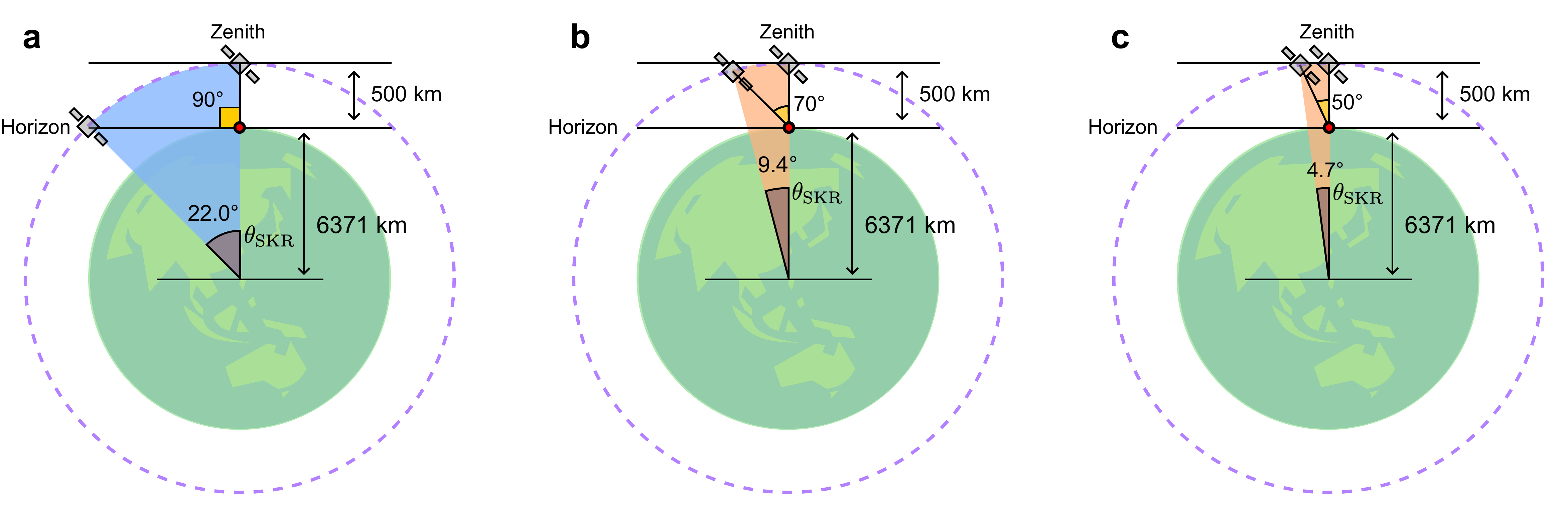}}
\caption{\label{fig:figure1} Satellite to ground model used in calculating the results presented in Fig.~6 of the main text. The satellite orbit is indicated by the purple dashed line and the red dot represents the optical ground station. \textbf{(a)} shows the geometric setting used to compute the duty cycle of our protocol. Regardless of weather conditions, the maximum zenith angle (shaded in yellow) where the secure key rate (SKR) exceeds $10^{-4}$ bits/use is $90^{\circ}$. Notably, our protocol achieves SKR delivery throughout the entire 3-hour window, which represents the maximum possible transmission window for any QKD protocol (CV-QKD or DV-QKD) in this scenario, resulting in a 100\% duty cycle. Figures \textbf{(b)} and \textbf{(c)} illustrate the model used to assess the duty cycle of a conventional GG02 protocol under good and bad weather conditions, respectively. In favorable weather, GG02 attains maximum zenith angles of approximately $70^{\circ}$, yielding a duty cycle of approximately 43\%. This drops to 21\% in adverse weather due to a reduced maximum zenith angle of $50^{\circ}$. The model incorporates the parameters outlined in Table~\ref{tab:free_space_parameters_table}. It is important to emphasize that the aspect ratio between element dimensions and lengths depicted in the figure are not precise. The representation has been adjusted to represent the model succinctly. As a consequence the zenith angles look inaccurate. However, the stated values are accurate and employed in the duty cycle calculations.}
\end{figure*}
For an optical ground state elevation of 0~km, the system can be visualized as illustrated in Fig.~\ref{fig:figure1}. A satellite in Low Earth Orbit (LEO), traveling at an average speed of 7.6 km/s~\cite{lu2022micius}, typically completes 15 orbits around the Earth in a single day. During this period, the satellite is visible from the ground station for approximately 3 hours. Consequently, 100\% duty cycle of secret key rate (SKR) delivery corresponds to 3 hours, irrespective of the tolerable minimum SKR. In the results shown in Fig.~6 of the main text, we assume the minimum tolerable secret key rate~(or threshold SKR) to be $10^{-4}$ bits per use. The duty cycle is determined by calculating the ratio of two arc lengths: the first from the zenith to the satellite at the altitude where the secret key rate exceeds the $10^{-4}$ bits/use threshold, and the second from the zenith to the horizon. Upon simplification, the duty cycle is given by
\begin{equation}
\text{Duty~cycle} = \frac{\theta_{\text{SKR}}}{{22}^{\circ}} \times 100\%,
\end{equation}
where the angle $\theta_{\text{SKR}}$ is the angle made between the satellite and the zenith with respect to the centre of the Earth. $\theta_{\text{SKR}}$ can be calculated from the elevation angle $\epsilon$~(=$~90^{\circ}-$  zenith angle) when the SKR crosses $10^{-4}$ bits per use and is given by 
\begin{equation}
\theta_{\text{SKR}} = [90 - \epsilon] - \text{sin}^{-1}\bigg[\cos[\epsilon]\frac{R_E}{R_E+L_{zen}}\bigg].
\end{equation}
\begin{table}[hbt!]
\caption{\label{tab:free_space_parameters_table}
Parameters of the Satellite-to-ground communication model in Fig.~6 of the main text.
}
\begin{tabular}{|c|c|}
\hhline{|=|=|}
Radius of the Earth, $R_E$ (km) & 6371 \\ \hline
Satellite Altitude at Zenith, $L_{zen}$ (km) & 500 \\ \hline
Optical Ground Station Elevation, $L_{OGS}$ (km) & 0 \\ \hline
Transmitter Aperture Diameter, $D_t$ (m) & 0.3 \\ \hline
Receiver Aperture Diameter, $D_r$ (m) &  1 \\ \hline
Transmitter Optics Efficiency, $T_t$ & 0.9 \\ \hline
Receiver Optics Efficiency, $T_r$ &  0.9 \\ \hline
Pointing Loss Efficiency, $L_p$ &  0.1 \\ \hline
Atmosphere Thickness, $L_{atm}$ (km) & 20 \\ \hline &\\[-4pt]
\shortstack{\vspace{1.5mm}Visibility, $V$ (km)} & \shortstack{\vspace{0.5mm}20 km for bad weather conditions \\
200 km for good weather conditions\vspace{0.5mm}}  \\[2pt] \hline &\\[-4pt]
\shortstack{\vspace{1.5mm}Refractive Index Structure Parameter, $C_n^2$ (m$^{-2/3}$)} & \shortstack{\vspace{0.5mm}$10^{-13}$ for bad weather conditions \\ \shortstack{$10^{-16}$ for good weather conditions}\vspace{0.5mm}} \\[2pt] \hline
Probability Threshold, $p_{th}$ & \shortstack{\vspace{-1mm}$10^{-6}$\vspace{0.5mm}} \\ \hline
Wavelength, $\lambda$ (nm) & 1550 \\ \hline
Modulation Variance, $V_{mod}$ (SNU) & 8 \\ \hline
Detector Efficiency, $\eta$ (\%) & 90\% \\ \hline
Detector Noise, $\epsilon_{det}$ (SNU) & 0.0135 \\ \hline
Excess Channel Noise, $\epsilon_{ch}$ (SNU) & 0.0186 \\ \hline
Detection Type & Homodyne \\ \hline
Reconciliation Efficiency, $\beta$ (\%) & 90\% \\
\hhline{|=|=|}
\end{tabular}
\end{table}

\FloatBarrier

\end{document}